\documentclass[11pt,a4paper]{article}
\usepackage{jheppub,amsmath,  amssymb,slashed,url,bm,textgreek,upgreek}
\usepackage{graphicx}
\usepackage{epstopdf}
\usepackage{esint}
\usepackage{braket}
\usepackage{appendix}
\usepackage{subcaption}

\definecolor{mypink}{RGB}{255,40,250} 

\hypersetup{
    colorlinks=true,
    linkcolor=blue,
    citecolor=mypink,
    urlcolor=blue
}

\title{Kaluza-Klein Gravitons in a Higher Curvature Warped Geometry : A New Perspective}

\author[a]{Abhirup Karmakar}
\author[a]{Isika Mandal}
\author[a]{Soumitra SenGupta}
\affiliation[a]{School of Physical Sciences, Indian Association for the Cultivation of Science, 2A $\&$ 2B, Raja S.C. Mullick Road Kolkata - 700032, India}
\emailAdd{abhirup04karmakar@gmail.com}
\emailAdd{isikamandal1@gmail.com}
\emailAdd{tpssg@iacs.res.in}
\abstract{Kaluza-Klein (KK) Gravitons are the direct collider imprints of the higher dimensional bulk physics in our four dimensional universe, arising from the compactification of an extra spatial dimension. In this work, we consider a two-brane warped geometry with a 5D $f(\mathcal R) = \mathcal R + \alpha\mathcal R^2$ gravity along with a cosmological constant $\Lambda$. The warped spacetime provides an elegant resolution of the gauge-hierarchy problem without introducing any intermediate scale, while the Planck-scale curvature of the underlying $AdS_5$ bulk naturally motivates the inclusion of higher-curvature corrections. For small values of the higher-curvature parameter ($\alpha$), we obtain the leading-order back-reacted warp factors perturbatively from the modified gravitational field equations. In the backdrop of a warped braneworld model, we have solved the Schr\"odinger-like equation governing the graviton fluctuations using a Euclidean path integral formalism, yielding the KK graviton spectrum and normalized wavefunctions directly from the corresponding quantum-mechanical propagator. Treating these results as the unperturbed background, we analytically determine the higher curvature corrections to the KK graviton spectrum and their couplings to Standard Model (SM) matter fields. We find that there is an appreciable upward shift in the KK graviton masses while leaving the graviton-SM couplings only mildly modified as compared to a model with only Einstein gravity in the bulk. However the net cross-section of processes involving virtual gravitons appears to be suppressed whereas the dilepton and diphoton decay widths of the gravitons are significantly enhanced because of the higher curvature corrections. Overall, these effects lead to observable modifications to both the production and decay signatures of massive KK gravitons and may be probed in some future precision collider experiments.
}

\begin{document}\maketitle

\section{Introduction}\label{intro}

The world of subatomic particles is best described by the Standard Model (SM) of elementary particles till today. The validity of SM has been confirmed with a great accuracy in various experiments up to TeV scale. The discovery of Higgs boson in the Large Hadron Collider (LHC) in 2012 indeed was a major breakthrough in this endeavor. Despite being a successful theory, SM encounters a long-standing but unresolved problem in the context of the stability of the mass of Higgs boson against a large radiative correction. This is known as the \textit{gauge hierarchy problem} or the \textit{fine-tuning problem}~\cite{1}. The two most popular models proposed in the context of this problem are - \textit{supersymmetry} and \textit{extra-dimensional} models~\cite{2,3,4,5,6,7,8,9,10,11,12,13,14,15,16,17,18,19,20,21,22,23}. However, interests continue to grow towards the extra-dimensional models due to the absence of any signature of supersymmetry near the TeV scale so far. On the other hand, the search for extra spatial dimensions has been an active research arena for a long time due to historical reasons. It started growing when Kaluza and Klein first proposed a possibility to unify gravity and electromagnetism by introducing an extra spatial dimension. Since then there has been a plentitude of work exploring this feature extensively in different contexts. For example: the origin of neutrino masses~\cite{24,25}, fermion mass hierarchy~\cite{26}, inflationary cosmology~\cite{27,28}, bouncing phenomena in cosmology~\cite{29,30,31}, galactic structure in astrophysics~\cite{32,33} and the one which is most famous, that is, resolution of the gauge-hierarchy problem or the fine-tuning problem of Higgs mass~\cite{34}. The concept of extra spatial dimensions come naturally in \textit{string theories} or some \textit{string-inspired models}~\cite{35}. These models can be broadly classified into two categories - with the large extra dimensional radii~\cite{2} and with the small extra dimensional radii~\cite{12}. In both cases, the extra dimensions are compactified under different topological configurations such that the effective 4D physics emerges from the higher dimensional theory and the imprints of the higher dimensional theory generally gets reflected in the effective 4D theory. Among these models, the 5D warped geometry model (with Einstein gravity in the bulk) proposed by Randall and Sundrum (RS)~\cite{12,13} assumed a special significance because firstly, it resolves the gauge hierarchy problem without introducing any other intermediate scale in the theory and secondly, the modulus of the extra dimension can be stabilized by introducing a bulk scalar field, as first proposed by Goldberger and Wise~\cite{36,37}, without any unnatural fine tuning of the model parameters. This two-brane RS model, along with its variants, have been thoroughly studied over the past two decades~\cite{38,39,40,41,42,43,44,45,46,47,48,49,50,51,52,53,54,55,56,57,58}. It is worth mentioning that a warped solution, though not exactly same as the RS model, (a throat-like geometry) can be found from string theory which as a fundamental theory predicts the inevitable existence of the extra spatial dimensions~\cite{59,60}. The RS model essentially captures the important features of this throat geometry in a simple way through the phenomenology of the \textit{Kaluza-Klein graviton modes}~\cite{61,62,63,75}. Kaluza-Klein (KK) graviton modes are the direct collider imprints of this warped extra dimensional theory and these massive KK modes couples with the standard model matter fields with a much larger coupling, at TeV scale and can be probed in LHC. The detectors in LHC are designed to explore these possible signatures of the warped extra dimensions through various decay channels of KK gravitons. While the CMS~\cite{64,65} detector searches the signal of the extra dimension through the final states of the decay into leptons, photons and hadrons, the ATLAS~\cite{66} detector is designed to capture the dileptonic decay of the KK gravitons.\\

\noindent In this work, we consider a two-brane 5D warped geometry with a higher-curvature $f(\mathcal R)$ bulk theory. $f(\mathcal R)$ theories of gravity arises from a simple generalization of the Einstein-Hilbert action from $\mathcal R$ (Ricci scalar) to some arbitrary function $f(\mathcal R)$~\cite{67,68}. Such a choice for the bulk emerges naturally from the requirement of setting a Planckian cosmological constant to address the gauge hierarchy problem. The consideration of the specific two-brane warped geometry is due to its robustness to resolve the gauge-hierarchy problem without bringing in any additional intermediate scale in the theory and as said earlier the inclusion of this bulk higher curvature term is no-doubt more realistic as the bulk spacetime in this model is endowed with a negative cosmological constant ($AdS_5$ geometry) of the order of Planck scale~\cite{69}. A new warped metric in the form of perturbative corrections to the original RS solution where the correction is generated due to higher curvature terms was first proposed in Ref.~\cite{70} by Elahi, Mandal and SenGupta. They also resolved the gauge-hierarchy problem as well as provided a novel mechanism to stabilize the modulus without introducing any external stabilizing field within this framework. The graviton KK spectrum in the RS model is conventionally obtained by solving the linearized gravitational field equations, which reduce to a Schr\"odinger-like eigenvalue problem for the extra-dimensional wavefunctions. In the present work, we revisit this problem from a different perspective by employing a Euclidean path-integral formalism. Rather than solving the differential equation directly, the graviton wavefunctions and mass spectrum are extracted from the corresponding quantum-mechanical propagator, thereby providing an alternative and complementary formulation of the RS KK graviton sector. Treating the obtained results as the perturbative background, we extend the perturbative framework developed by Elahi \textit{et al.} and find the KK graviton modes in this model with a stabilized modulus. In other words, we find the perturbative corrections to KK graviton wavefunctions, spectrum as well as the coupling of KK gravitons with the SM matter fields at the TeV brane as compared to the original RS model, where the corrections are generated due to higher curvature effects in the bulk. We find an appreciable upward shift in the KK graviton masses, while the graviton-SM couplings remain only mildly modified, leading to suppressed virtual graviton production rates and enhanced dilepton and diphoton decay signatures that may be probed in future precision collider experiments. Finally, we give a validity of our obtained results using the naive dimensional analysis and show the convergence of effective field theory (EFT) for the allowed parameter space.\\

\noindent The paper organizes as follows: In section \ref{framework}, we lay out the framework of our model in four subsections. In sections \ref{action} and \ref{perturb}, we set up the notation and derive the back-reacted metric perturbatively from the modified field equations. In sections \ref{grav_coupling} and \ref{phys_mass}, we discuss the relation between the 5D and 4D gravitational strengths and a possible resolution of the gauge-hierarchy problem in this setup, with a brief mention of modulus stabilization. In section \ref{review}, we review the RS graviton problem in pure Einstein gravity and derive the KK graviton spectrum and wavefunctions; a detailed Euclidean path-integral re-derivation is given in Appendix \ref{A}. Section \ref{modified} contains the main results of our work, where we solve the KK graviton problem for the higher-curvature back-reacted model within the perturbative framework. This section has five subsections: in section \ref{intro_modified}, we set up the graviton equations and the perturbative scheme; in sections \ref{analysis1} and \ref{analysis2}, we present two independent approaches that lead to the same KK graviton results for the higher-curvature scenario, with some coefficients given in Appendix \ref{B}. In section \ref{coupling}, we discuss the coupling of the KK modes with SM matter fields on the TeV brane, and in section \ref{pheno}, we analyze the phenomenological implications of the higher-curvature corrections. In section \ref{EFT}, we provide the naive dimensional analysis and show the convergence of EFT for allowed parameter space. Finally, in section \ref{discussion}, we summarize our findings.

\section{Framework: 5D Warped Geometry}\label{framework}

\subsection{Action and Field Equations}\label{action}

We briefly review the background metric discovered by Randall and Sundrum to set up our notations. The whole set up is a 5D Einstein gravity with a bulk cosmological constant $\Lambda$ with two 3-branes located in the orbifold $S^1/\mathbb Z_2$ at the extra dimensional coordinate $z = 0$ (Planck brane or hidden brane) and $z = r_c$ (TeV brane or visible brane). The full 5D gravitational action
\begin{equation}\label{2.1}
S = \int d^5x\sqrt{-g_5}\;[M_*^3\mathcal R - \Lambda] - \mathcal T_{hid}\int d^4x \sqrt{-g_{(+)}} - \mathcal T_{vis}\int d^4x \sqrt{-g_{(-)}}
\end{equation}
where, $M_*$ is the 5D Planck mass scale defined in term of 5D graviational constant $G_5$ as $M_*^{-3} = 16\pi G_5$. $\mathcal T_{hid,vis}$ are the brane tensions of hidden brane and of visible brane. $\mathcal R$ is the 5D Ricci scalar. The induced metric on positive tension brane is $g_{(+)\mu\nu} = g_{\mu\nu}$ and on negative tension brane is $g_{(-)\mu\nu} = e^{-2kr_cT}g_{\mu\nu}$. Note the range of the extra dimension in $S^1/\mathbb Z_2$ is $z \in [-r_c,r_c]$ with $z$ and $-z$ identified. Randall and Sundrum showed that there exists a metric solution which respects the 4D Poincar\'e invariance
\begin{equation}\label{2.2}
    ds^2 = e^{-2kT|z|}\eta_{\mu\nu}dx^\mu dx^\nu + T^2dz^2
\end{equation}
if and only if the following \textit{fine-tuning conditions} are satisfied
\begin{equation}\label{2.3}
    \mathcal T_{hid} = -\mathcal T_{vis} = - \frac{\Lambda}{k}\;\;\;\;;\;\;\;\; \Lambda = -12M_*^3k^2
\end{equation}
where, $\eta_{\mu\nu}$ is the flat Minkowski metric and $k$ is the $AdS_5$ curvature scale. Throughout the paper, we will be using the mostly plus $(-+++)$ metric convention.\\

\noindent In the original RS model, the value of the bulk cosmological constant was chosen to be of the order of the Planck scale. This naturally motivates us to consider a small higher curvature correction to the bulk Einstein gravity. Hence we consider the following bulk gravitational action
\begin{equation}\label{2.4}
S = M_*^3\int d^5x\sqrt{-g_5}\;f(\mathcal R) - \mathcal T_{hid}\int d^4x \sqrt{-g_{(+)}} - \mathcal T_{vis}\int d^4x \sqrt{-g_{(-)}}
\end{equation}
where, $f(\mathcal R) = \mathcal R + \alpha\mathcal R^2 - \Lambda$ with $\alpha$ being mass dimension $-2$. $\alpha$ corresponds to the small higher curvature correction to the Einstein theory which will be our perturbative parameter in the subsequent discussions for this section. From now on, we are setting $M_* = 1$ for the ease of calculations.\\

\noindent We are interested to study the warped solutions of the above bulk $f(\mathcal R)$ gravity. This inspires us to take the following metric ansatz
\begin{equation}\label{2.5}
    ds^2 = e^{-2A(y)}\eta_{\mu\nu}dx^\mu dx^\nu + B^2(y)dy^2
\end{equation}
where, $y \equiv Tz$. The field equations for a general $f(\mathcal R)$ theory~\footnote{In this paper, we are considering $f(\mathcal R)$ theories in the metric formalism only} are the following
\begin{equation}\label{2.6}
    f_{\mathcal R}(\mathcal{R})\mathcal R_{AB} - \frac{1}{2}f(\mathcal{R})g_{AB} - \left(\nabla_A\nabla_B - g_{AB}\Box\right)f_{\mathcal R}(\mathcal R) = \frac{1}{2}T_{AB}
\end{equation}
where, $A,B = 0,1,2,3,4$ are the 5D indices with $4$ denoting the extra spatial coordinate and $f_{\mathcal R}(\mathcal R)$ denotes the derivative of $f(\mathcal R)$ with respect to $\mathcal R$. $T_{AB}$ is the energy-momentum tensor corresponding to some matter action $\mathcal S_M$
\begin{equation}\label{2.7}
    T_{AB} = -\frac{2}{\sqrt{-g_5}}\frac{\delta \mathcal S_M}{\delta g^{AB}}
\end{equation}
But we are not considering any matter for our present context, so we take $T_{AB} = 0$. Some of the important components of the Christoffel symbols, Ricci tensor and the Ricci scalar for the metric ansatz in eq.(\ref{2.5}) are
\begin{align}\label{2.8}
\Gamma^4_{44} = \frac{B'(y)}{B(y)}\;\;\;\;\;;\;\;\;\;\; \Gamma^4_{\mu\nu} = \eta_{\mu\nu}e^{-2A(y)}\frac{A'(y)}{B^2(y)}\;\;\;\;\;;\;\;\;\;\;\Gamma^\mu_{4\nu} = \delta^\mu_\nu A'(y)
\end{align}
\begin{align}\label{2.9}
&\mathcal R_{\mu\nu} = \frac{e^{-2A(y)}}{B^3(y)}\eta_{\mu\nu}\left[B(y)\Big\{A''(y) - 4A'^2(y)\Big\} -A'(y)B'(y)\right] \nonumber \\
&\mathcal R_{44} = -4\left[A'^2(y) - A''(y) + \frac{A'(y)B'(y)}{B(y)}\right] \nonumber \\
&\mathcal R = -\frac{4}{B^3(y)}\left[2A'(y)B'(y) + B(y)\Big\{5A'^2(y) - 2A''(y)\Big\}\right]
\end{align}
where, prime denotes the derivative with respect to the extra spatial coordinate $y$ and $\mu,\nu = 0,1,2,3$ are the usual 4D indices. In the next subsection, we will solve the modified field equations in (\ref{2.6}) perturbatively to find the solutions for $A(y)$ and $B(y)$ up to leading order in the higher curvature parameter $\alpha$.

\subsection{Perturbative Solutions of the Field Equations}\label{perturb}

We are now proposing to solve the modified field equations in (\ref{2.6}) for $f(\mathcal R) = \mathcal R + \alpha\mathcal R^2 - \Lambda$ (note that, now $\alpha$ is dimensionless as we set $M_*=1$). We consider $\alpha$ is small (that is, $0<\alpha < 1$) such that $\mathcal R^2$ in $f(\mathcal R)$ is a small correction over $\mathcal R$. Therefore we consider the following ansatz for $A(y)$ and $B(y)$~\cite{70}
\begin{equation}\label{2.10}
A(y) = ky + \alpha A_1(y)\;\;\;\;\;;\;\;\;\;\; B(y) = 1 + \alpha B_1(y)
\end{equation}
where, $ky$ denotes the unperturbed RS-like part of the warp factor $A(y)$ and $A_1(y),B_1(y)$ are the first order perturbative corrections to $A(y)$ and $B(y)$, respectively. We use the metric ansatz (\ref{2.5}) and the ansatz for $A(y),B(y)$ in (\ref{2.10}) in the modified field equations (\ref{2.6}). The $\mu\nu$ component of the field equations reduces to the following equation in the first order approximation in $\alpha$
\begin{equation}\label{2.11}
\left(6k^2 + \frac{\Lambda}{2}\right) + \alpha\left[-40k^4 - 3A_1''(y) + 12kA_1'(y) + 3kB_1'(y) -12k^2B_1(y) - (\Lambda + 12k^2)A_1(y)\right] = 0
\end{equation}
Individually equating the coefficients of the $\mathcal O(1)$ and $\mathcal O(\alpha)$ terms to zero, we get
\begin{equation}\label{2.12}
    \Lambda = -12k^2
\end{equation}
and
\begin{equation}\label{2.13}
    -40k^4 - 3A_1''(y) + 12kA_1'(y) + 3kB_1'(y) -12k^2B_1(y) = 0
\end{equation}
Eq.(\ref{2.12}) tells us immediately that in order to have a valid warped solution, the bulk should be Anti de-Sitter (that is, $\Lambda < 0$) as $k > 0$ always. Now similarly the $44$ component of the field equation reduces to
\begin{equation}\label{2.14}
    -40k^4 + 12kA_1'(y) - 12k^2B_1(y) = 0
\end{equation}
Solving the equations (\ref{2.13}) and (\ref{2.14}), we get the following solution for $A(y)$ and $B(y)$
\begin{align}\label{2.15}
A(y) = ky + \frac{10}{3}\alpha k^3y + \frac{1}{2}\alpha b_0ky^2\;\;\;\;\;;\;\;\;\;\; B(y) = 1 + \alpha b_0y
\end{align}
In principle, one can consider higher order corrections as well, but since we want to explore the small $\alpha$ regimes, we restrict ourselves only up to the first order. One can obtain the modified fine-tuning conditions up to the leading order in $\alpha$
\begin{equation}\label{2.16}
    \mathcal T_{hid} = - \mathcal T_{vis} = 12k + 40\alpha k^3
\end{equation}
Note that $\alpha$ is the higher curvature parameter, whereas $b_0$ is the back-reaction parameter. $\alpha,b_0$ are the free parameters of our constructed model. Now we will be discussing the relation between the fundamental gravitational strength and the 4D effective strength for our back-reacted model.

\subsection{Gravitational Couplings}\label{grav_coupling}

In order to determine the strength of gravitational interaction, we reintroduce $M_*$ and then the 5D gravitational action will contain the following piece
\begin{equation}\label{2.17}
    \mathcal S \supset M_*^3\int d^5x\sqrt{-g_5}\; \mathcal R
\end{equation}
where, $\mathcal R$ is the 5D Ricci scalar which contains the 4D Ricci scalar $R$ as $\mathcal R \supset e^{2A(y)}R$. This in turn reduces the 5D action into a 4D effective action as follows
\begin{equation}\label{2.18}
    \mathcal S_{eff} \supset \underbrace{M_*^3\int_0^b dy\;B(y)e^{-2A(y)}}_{M_{Pl}^2}\;\int d^4x\sqrt{-g_4}\;R
\end{equation}
where, we immediately identify the 4D Planck scale $M_{Pl}^2$ as
\begin{equation}\label{2.19}
    M_{Pl}^2 = M_*^3\int_0^b dy\;B(y)e^{-2A(y)}
\end{equation}
Using eq.(\ref{2.15}), we can explicitly evaluate the above integral and find the relation between 4D and 5D Planck scales as follows
\begin{align}\label{2.20}
M_{Pl}^2 = M_*^3\Bigg[\frac{1}{2k}&\left(1 - e^{-2Kb - \varepsilon kb^2}\right) \nonumber \\
&+ \frac{1}{2}\sqrt{\frac{\pi}{\varepsilon k}}\left(1 - \frac{K}{k}\right)e^{K^2/\varepsilon k}\Big\{\mathbf{erf}\left(\sqrt{\varepsilon k}\;b + \frac{K}{\sqrt{\varepsilon k}}\right) - \mathbf{erf}\left(\frac{K}{\sqrt{\varepsilon k}}\right)\Big\}\Bigg]
\end{align}
where, $K$ and $\varepsilon$ are defined to be the following quantities
\begin{equation}\label{2.21}
    K = k + \frac{10}{3}\alpha k^3 \;\;\;\;;\;\;\;\; \varepsilon = \alpha b_0
\end{equation}
and $\mathbf{erf}(x)$ is the \textit{error function} defined as
\begin{equation}\label{2.22}
    \mathbf{erf}(x) = \frac{2}{\sqrt{\pi}}\int_0^xdu\;e^{-u^2}
\end{equation}
For $\alpha \to 0$ we get back the standard RS relation between $M_{Pl}$ and $M_*$
\begin{equation}\label{2.23}
    M_{Pl}^2 = \frac{M_*^3}{2k}\left(1 - e^{-2kb}\right)
\end{equation}

\noindent In the next subsection, we will briefly discuss the physical Higgs mass scale and the modulus stabilization in this higher curvature scenario.

\subsection{Physical Mass Scale and Modulus Stabilization}\label{phys_mass}

The matter fields on the TeV brane couple to the induced metric $g_{\mu\nu(-)} = e^{-2A(b)}g_{\mu\nu}$. The action for the Higgs field $\mathcal H$ can be written as (setting $M_* = 1$ again)
\begin{equation}\label{2.24}
    \mathcal S_{Higgs} = \int d^4x\sqrt{-g_{(-)}}\left[g^{\mu\nu}_{(-)}\mathcal D_\mu\mathcal H^\dagger\mathcal D_\nu\mathcal H - \lambda\left(\mathcal H^\dagger\mathcal H - v^2\right)^2\right]
\end{equation}
We canonically normalize the Higgs field as $e^{-A(b)}\mathcal H$, then
\begin{equation}\label{2.25}
    \mathcal S_{Higgs} = \int d^4x\sqrt{-g}\left[g^{\mu\nu}\mathcal D_\mu\mathcal H^\dagger\mathcal D_\nu\mathcal H - \lambda\left(\mathcal H^\dagger\mathcal H - e^{-2A(b)}v^2\right)^2\right]
\end{equation}
Therefore, the physical mass scale on the TeV brane is given by
\begin{equation}\label{2.26}
    v_{phy} \equiv  e^{-A(b)}v
\end{equation}
We know that $v \sim M_{Pl}$, so to generate TeV scale ($v_{phy}\sim$ TeV) from $M_{Pl}$, we only require $A(b) \sim 36$. This can be achieved for $b_0 \sim \mathcal O(1-10)$, $\alpha\sim\mathcal O(0.1)$ and $b \sim\mathcal O(1-10)$. We do not require any large hierarchies between the fundamental parameters $\alpha,\;b_0,\;k,\;b$. Hence for this model, it is possible to keep the scale of gravity unchanged while the scale of Higgs gets exponentially suppressed. \\

\noindent For the higher curvature scenario, we can stabilize the modulus field, namely the $T(x)$ field. This modulus stabilization mechanism in higher dimensional $f(\mathcal{R})$ gravity has been first proposed in Ref.~\cite{70}. We need to have $0<\alpha<1$ and $b_0>0$ in order to obtain modulus stabilization in ghost-free quadratic $f(\mathcal R)$ gravity (thus, $f'(\mathcal R) > 0$ as well as $f''(\mathcal R) > 0$). But for $\alpha\to 0$ stabilization no longer holds. Since the modulus stabilization in any braneworld model is an important requirement to make the prediction of the model more robust, it is worthwhile to consider the model in its stabilized version.\\

\noindent In the following section, we will be briefly reviewing the KK gravitons in usual Randall-Sundrum (RS) set up (Einstein gravity in the bulk).

\section{Review: KK Gravitons in Einstein Gravity}\label{review}

In this section, we will briefly review the Kaluza-Klein (KK) graviton phenomenology for pure Einstein gravity in the bulk described by the action (\ref{2.1}). For the discussion of KK graviton modes, we generally work in a different coordinate $\xi$ of the extra spatial dimension (also called the conformal coordinate), which is related to the $y$ coordinate as $d\xi = e^{ky}dy$. We will be working with any of the three coordinate choices for the extra spatial dimension $y$ or $z$ or $\xi$ depending on their usefulness in the calculations.\\

\noindent In order to find the KK expansion of the graviton modes, we go to the conformal frame of the metric in (\ref{2.2}) and parametrize the graviton fluctuations by
\begin{equation}\label{3.1}
    ds^2 = e^{-2A(\xi)}\left[\big\{\eta_{\mu\nu} + h_{\mu\nu}(x,\xi)\big\}dx^\mu dx^\nu + d\xi^2\right] \;\;\;\;;\;\;\;\; A(y) = \ln(k|\xi| + 1)\;\;,\;\;B(y) = 1
\end{equation}
After rescaling the tensor perturbation as $h_{\mu\nu}(x,\xi) = e^{\frac{3}{2}A(\xi)}\tilde h_{\mu\nu}(x,\xi)$, one can obtain the linearized version of Einstein's equation for the metric in (\ref{3.1}) as follows~\cite{56}
\begin{equation}\label{3.2}
    -\frac{1}{2}\partial^M\partial_M\tilde h_{\mu\nu} + \left[\frac{9}{8}\partial^MA\partial_MA - \frac{3}{4}\partial^M\partial_MA\right]\tilde h_{\mu\nu} = 0
\end{equation}
where, we are working in RS gauge choice (basically transverse-traceless gauge), that is, $h^\mu_{\;\mu} = 0$ and $\partial_\mu h^{\mu\nu} = 0$. Now we decompose the graviton fluctuations in KK modes (the standard method of separation of variables) as
\begin{equation}\label{3.3}
    \tilde h_{\mu\nu}(x,\xi) = \hat h_{\mu\nu}(x)\Psi(\xi)
\end{equation}
and we demand that $\hat h_{\mu\nu}$ to be a 4D mass eigenstate mode as $\Box \hat h_{\mu\nu} = m^2\hat h_{\mu\nu}$. Therefore, finally we get a Schr\"odinger-like equation for graviton KK modes
\begin{equation}\label{3.4}
    \left[-\partial_\xi^2 + V(\xi)\right]\Psi(\xi) = m^2\Psi(\xi)
\end{equation}
where,
\begin{equation}\label{3.5}
    V(\xi) = \frac{15}{4}\frac{k^2}{(k|\xi| + 1)^2} - \frac{3k\delta(\xi)}{k|\xi| + 1}
\end{equation}
which is the well-known \textit{volcano potential}. $\Psi$ is called the \textit{graviton wavefunction}. Here one can define some operators $Q = \partial_\xi + \frac{3}{2}A'$ and $Q^\dagger = -\partial_\xi + \frac{3}{2}A'$ and can rewrite eq.(\ref{3.4}) as $Q^\dagger Q\Psi = m^2\Psi$, which is analogous to a supersymmetric quantum mechanics problem. This way of writing ensures the positivity of the eigenvalues $m^2 > 0$ and one can extract the zero mode ($m=0$) solution very easily from $Q\Psi = 0$. Therefore, the zero mode solution is given by
\begin{equation}\label{3.6}
    \Psi(\xi) = \frac{\sqrt{2k}}{(k|\xi| + 1)^{3/2}}
\end{equation}
where, the factor $\sqrt{2k}$ in the numerator is the normalization we obtained considering the two-brane separation is sufficiently large compared to the fundamental mass scale. This approximation is certainly valid to take as for the case of compactified extra dimension, the zero mode never decouples (contrary to the flat extra dimensional models) when we consider the extra dimension to be sufficiently large. Having a zero mode solution is completely natural as we have not broken the 4D Poincar\'e invariance, so we expect one massless 4D graviton to exist. The nontrivial wavefunction for the graviton zero mode peaks around the Planck brane and decays far away having a small tail. This suggests that the gravity itself is actually \textit{localized} around the Planck brane. One can show this localization of gravity by calculating the semi-classical WKB tunneling amplitude of the zero mode from Planck brane
\begin{equation}\label{3.7}
    T \sim \exp\left(-2\int_{\xi_{min}}^{\xi_{max}}\sqrt{V(\xi) - m^2}\right) \xrightarrow{m\to 0} 0
\end{equation}
where, $\xi_{min},\xi_{max}$ are the classical turning points in the volcano potential. See that for zero mode actually the tunneling probability tends to zero, indicating the localization of gravity at the Planck brane.\\

\noindent Now for calculating the massive KK graviton modes, we need to impose appropriate boundary conditions while solving the eq.(\ref{3.4}) which are needed for the $S^1/\mathbb Z_2$ orbifold projection $y \to -y$ as follows
\begin{align}\label{3.8}
\partial_\xi\Psi(\xi) &= -\frac{3}{2}k\Psi(\xi)\Bigg|_{\xi=0} \nonumber \\
\partial_\xi\Psi(\xi) &= -\frac{3}{2}\frac{k}{k|\xi_0| + 1}\Psi(\xi)\Bigg|_{\xi=\xi_0}
\end{align}
where, $\xi_0 \equiv (e^{kb} - 1)/k$ denoting the separation of the two branes in $\xi$ coordinate and $b \equiv r_cT$ is the brane separation in $y$ coordinate. The graviton equation in (\ref{3.4}) subject to the boundary conditions in eq.(\ref{3.8}) can be solved nicely by evaluating a Euclidean path integral as follows
\begin{equation}\label{3.9}
\mathcal K_E(\rho_f, \rho_i; T) = \int \mathcal{D}[\rho(\tau)] \exp\left( -\int_0^T d\tau \left[ \frac{1}{4}\left(\frac{d\rho}{d\tau}\right)^2 + \frac{15}{4\rho^2} \right] \right)
\end{equation}
where, $\rho \equiv \xi + 1/k$ for $\xi > 0$. Evaluating the path integral systematically, one can find the normalized solution for the massive modes, which can be expressed in terms of Bessel's function as follows
\begin{equation}\label{3.10}
\Psi_n(\xi) = \frac{\sqrt{2k}\;e^{-kb}}{J_2(x_n)}(k\xi+1)^{1/2}J_2\left(m_{n}(\xi + 1/k)\right)\;\;\;\;\;;\;\;\;n = 1,2,3,...
\end{equation}
where, $x_n$'s are the zeros of the Bessel function $J_1$, that is, $J_1(x_n) = 0$. And $m_{n}$ is the spectrum of the full massive KK tower of gravitons which is completely determined by the two boundary conditions in eq.(\ref{3.8})
\begin{equation}\label{3.11}
    m_{n} = kx_ne^{-kb} \sim \mathcal O(\text{TeV})
\end{equation}
The massive KK modes are spaced by TeV suggests that it is better to think RS model with the extra dimension of size $1/$TeV, that is a relatively large extra dimension. A more detailed derivation of the above results using the path integral approach can be found in Appendix \ref{A}.\\

\noindent In the following section, we are going to discuss the perturbative modifications in graviton KK modes for small higher curvature effects along with the modified couplings between the graviton and SM fields which may have serious phenomenological or experimental implications.

\section{KK Gravitons in Higher Curvature Back-Reacted Gravity}\label{modified}

\subsection{Introduction}\label{intro_modified}

In section \ref{review}, we discussed about the discrete KK graviton spectrum arising due to the compactification of extra spatial dimension with the bulk theory to be pure Einstein (described by action in (\ref{2.1})). Having established the background details of the topic, we now investigate what will going to happen if we consider small higher curvature effects, namely how the KK graviton spectrum, graviton wavefunctions and couplings going to change. \\

\noindent To discuss the graviton fluctuation in this higher curvature scenario, we need to parametrize the metric in (\ref{2.5}) as follows
\begin{equation}\label{4.1}
    ds^2 = e^{-2A(y)}\left[\eta_{\mu\nu} + h_{\mu\nu}(x,y)\right]dx^\mu dx^\nu + B^2(y)dy^2
\end{equation}
Linearizing the $f(\mathcal R)$ gravity field equations in (\ref{2.6}) for the metric in (\ref{4.1}) using the RS gauge choice $h^\mu_{\;\mu} = 0$ and $\partial_\mu h^{\mu\nu} = 0$, we get the following linearized field equations
\begin{equation}\label{4.2}
\bar h''_{\mu\nu} + \left(\frac{f_\mathcal R'}{f_\mathcal R} - 4A' - \frac{B'}{B}\right)\bar h'_{\mu\nu} + B^2e^{2A}\Box\bar h_{\mu\nu} = 0
\end{equation}
where, $\bar h_{\mu\nu} = e^{-2A(y)}h_{\mu\nu}$ and the prime denotes derivative with respect to extra spatial coordinate $y$. Again decomposing $\bar h_{\mu\nu}$ as $\bar h_{\mu\nu} = \hat h_{\mu\nu}(x)\Psi(y)$ with $\hat h_{\mu\nu}(x)$ to be the mass eigenstate mode $\Box\hat h_{\mu\nu} = m^2\hat h_{\mu\nu}$, we get the equation for graviton wavefunction $\Psi(y)$ as follows
\begin{equation}\label{4.3}
    \left[\partial_y^2 + \left(\frac{f'_\mathcal R}{f_\mathcal R} - 4A' - \frac{B'}{B}\right)\partial_y + B^2e^{2A}m^2\right]\Psi(y) = 0
\end{equation}
where, the graviton wavefunctions $\Psi(y)$'s obey the following orthonormality condition
\begin{equation}\label{4.4}
    \int_0^b dy\;B(y)e^{-2A(y)}\Psi_n(y)\Psi_m(y) = \delta_{nm}
\end{equation}
Now we rewrite the solutions of $A(y)$ and $B(y)$ from eq.(\ref{2.15}) as follows
\begin{equation}\label{4.5}
    A(y) = Ky + \frac{1}{2}\varepsilon ky^2 
    \;\;\;\;;\;\;\;\;B(y) = 1 + \varepsilon y 
\end{equation}
where, $K$ and $\varepsilon$ are defined as
\begin{equation}\label{4.6}
    K = k + \frac{10}{3}\alpha k^3 \;\;\;\;;\;\;\;\; \varepsilon = \alpha b_0
\end{equation}
We consider the shifted RS background $A(y) = Ky$ to be our perturbation background and $\varepsilon$ to be the perturbation parameter. Inserting the solutions of $A(y)$ and $B(y)$ in eq.(\ref{4.3}) from eq.(\ref{2.15}) for $f(\mathcal R) = \mathcal R + \alpha \mathcal R^2 - \Lambda$, the equation for the graviton wavefunction $\Psi(y)$ reduces to the following form
\begin{equation}\label{4.7}
    \partial_y^2\Psi(y) + P(y)\partial_y\Psi(y) + Q(y)\Psi(y) = 0
\end{equation}
with
\begin{equation}\label{4.8}
P(y) = - 4K - 4\varepsilon k y - \varepsilon \;\;\;\;;\;\;\;\; Q(y) = (1 + 2\varepsilon y + \varepsilon ky^2)e^{2 Ky}m^2
\end{equation}
Doing the following field redefinition
\begin{equation}\label{4.9}
    \Psi(y) = \exp\left(-\frac{1}{2}\int^yds\;P(s)\right)\zeta(y)
\end{equation}
we reduce the graviton equation in a more suitable form to work with
\begin{equation}\label{4.10}
    \left[-\partial_y^2 + \mathcal V(y)\right]\zeta(y) = 0
\end{equation}
where,
\begin{equation}\label{4.11}
\mathcal V(y) = \left(4K^2 - e^{2Ky}m^2\right) - \varepsilon\Big\{e^{2Ky}m^2\left(2y + ky^2\right) - 8Kky + (2k-2K)\Big\}
\end{equation}
From the knowledge of section \ref{review}, we exactly know the unperturbed solutions (for $\alpha = 0$) of the above differential equation in (\ref{4.7}). We rewrite those unperturbed wavefunctions from eq.(\ref{3.6}) and (\ref{3.9}), but for our shifted unperturbed background $A(y) = Ky$ ($\varepsilon = 0$, but $\alpha\neq 0$)
\begin{align}\label{4.12}
\zeta_0^{(0)}(y) &= \sqrt{2K}\;e^{-2Ky} \nonumber \\
\zeta_n^{(0)}(y) &= \frac{\sqrt{2K}\;e^{-Kb}}{J_2(x_n)} J_2\left(\frac{m_{n}}{K}e^{Ky}\right)\;\;\;\;\;;\;\;\;n =1,2,3,...
\end{align}
Here the superscript $^{(0)}$ denotes the unperturbed quantities in this section. Below we will provide the perturbative analysis of the higher curvature back-reacted KK graviton problem using two methods - firstly we will try to find the approximate solutions in closed form so that we can use the approximate relations in the phenomenological discussions. And secondly, we will provide an alternative method to find the same solutions using perturbation theory analogous to quantum mechanics. Although for the latter, it is difficult to obtain solutions in a closed form for some cases.

\subsection{Analysis I: Approximate Analytic Solutions}\label{analysis1}

In this section, we are going to present the approximate solutions for zero graviton mode and massive graviton modes for our model described by action in (\ref{2.4}). We will make approximations at the leading order of $\varepsilon$ in order to obtain a closed form solutions of the wavefunctions, which we will be using in the discussion of phenomenology with our constructed model in section \ref{pheno}.

\subsubsection{The Zero Graviton Mode}

As we have discussed in section \ref{review}, the graviton zero mode is essential for recovering 4D gravity on the branes. Any general gravity theory which respects the 4D Poincar\'e invariance should have a graviton zero mode. Now in particular for our case, $f(\mathcal R) = \mathcal R + \alpha\mathcal R^2 - \Lambda$ theory satisfies the 4D Poincar\'e invariance without any doubt. So we must get a graviton zero mode or in other words, we should get a higher curvature correction in the zero mode wavefunction as compared to the unperturbed one in eq.(\ref{3.6}). \\

\noindent To find zero mode, we set $m = 0$ at the onset in eq.(\ref{4.7}) and we get the following
\begin{equation}\label{4.13}
\left[-\partial_y^2 + \big\{4K^2 +\varepsilon F_1(y)\big\}\right]\zeta(y) = 0 \;\;\;\;;\;\;\;\; F_1(y) = 8Kky + (2K-2k)
\end{equation}
The above equation can be solved perturbatively and we find the following solution
\begin{equation}\label{4.14}
\zeta_{0}(y) = N_0\sqrt{2K}\,e^{-2Ky}
\left[1 + \varepsilon\left(-ky^2 - \frac{1}{2}y\right)\right]
\end{equation}
where, $N_0$ is a normalization constant for the zero mode wavefunction which we need to determine. Now the above equation can be written as
\begin{equation}\label{4.15}
    \zeta_{0}(y) = N_0\sqrt{2K}\exp\left(-2Ky - \frac{1}{2}\varepsilon y - \varepsilon ky^2\right)
\end{equation}
as $\varepsilon <1$ is the perturbative parameter. Using the field redefinition relation between $\zeta(y)$ and $\Psi(y)$ in eq.(\ref{4.9}), we get
\begin{equation}\label{4.16}
    \Psi_{0}(y) = N_0\sqrt{2K} 
\end{equation}
Using the eq.(\ref{2.19}) and the orthonormality condition in eq.(\ref{4.4}), we find the normalization constant $N_0 = \sqrt{M_*^3/2KM_{Pl}^2}$ and finally we express the normalized zero mode solution as
\begin{equation}\label{4.17}
    \Psi_0(y) = \frac{M_*^{3/2}}{M_{Pl}}
\end{equation}
which is essentially a constant. Note that this way of deriving the zero mode actually provides us the \textit{exact} zero mode solution, rather than a perturbative solution only. Now we will try to find the massive mode wavefunctions and the corresponding spectrum in the following subsection.

\subsubsection{The Massive Graviton Modes}

For finding the massive modes, we will consider the full equation in (\ref{4.3}). Now for our case $f(\mathcal R) = \mathcal R + \alpha\mathcal R^2 - \Lambda$ and using eq.(\ref{2.9}), we can show that $2\alpha\mathcal R' \simeq -\frac{200k^4}{3b_0^2}\varepsilon^3 \implies f_\mathcal R' / f_\mathcal R \sim \mathcal O(\varepsilon^3)$. So we can safely ignore $f_\mathcal R' / f_\mathcal R$ term for a first order analysis. Hence we left with the following equation
\begin{equation}\label{4.18}
    \Psi''(y) + \left(-4A' - \frac{B'}{B}\right)\Psi'(y) + B^2e^{2A}m^2\Psi(y) = 0
\end{equation}
We do a coordinate change $du = B(y)dy$ in order to get rid of the $B(y)$ factors in the above equation
\begin{equation}\label{4.19}
    \ddot\Psi -4\dot A\dot\Psi + e^{2A}m^2\Psi = 0
\end{equation}
where, dot denotes derivative with respect to the $u$ coordinate. Again we do another coordinate transformation as $\chi = \frac{m}{K}e^{A(u)}$ and we redefine the wavefunction as $\tilde\Psi(u) = e^{-2A(u)}\Psi(u)$. Then we get
\begin{equation}\label{4.20}
    \chi^2\tilde\Psi_{\chi\chi} + \chi\left(1 + \frac{\ddot A}{\dot A^2}\right)\tilde\Psi_\chi + \left(\frac{K^2}{\dot A^2}\chi^2 + \frac{2\ddot A}{\dot A^2} - 4\right)\tilde\Psi = 0
\end{equation}
where, the subscript $_{\chi}$ denotes the derivative with respect to $\chi$. Now we express the $u$ coordinate as a function of $y$ as follows
\begin{equation}\label{4.21}
    u(y) = \int^yd\eta\;B(\eta) = y + \frac{1}{2}\varepsilon y^2
\end{equation}
We now perturbatively invert the above relation to express $y = y(u)$ as
\begin{equation}\label{4.22}
    y(u) = u -\frac{1}{2}\varepsilon u^2 + \mathcal{O}(\varepsilon^2)
\end{equation}
Therefore, we express the warp factor $A(y)$ as a function of $u$ coordinate as follows
\begin{equation}\label{4.23}
    A(y) = Ky + \frac{1}{2}\varepsilon ky^2 \implies A(u) = Ku + \frac{1}{2}\varepsilon(k-K)u^2 + \mathcal{O}(\varepsilon^2)
\end{equation}
The quadratic piece is very small term compared to the linear piece, hence we can safely ignore that and we get $A(u)\simeq Ku$. So in $u$ coordinate, the problem approximately reduced to a RS-like problem and we can easily solve the equation below
\begin{equation}\label{4.24}
    \chi^2\tilde\Psi_{\chi\chi} + \chi\tilde\Psi_\chi + \left(\chi^2 - 2^2\right)\tilde\Psi = 0
\end{equation}
This is a standard Bessel differential equation, which has a general solution of the form
\begin{equation}\label{4.25}
    \Psi_n(u) \simeq N_ne^{2A(u)}\left[J_2\left(\frac{m_n}{K}e^{A(u)}\right) + \gamma_n Y_2\left(\frac{m_n}{K}e^{A(u)}\right)\right]
\end{equation}
where, $N_n$ is a normalization constant for the wavefunction $\Psi_n$. $J_2,Y_2$ are the Bessel and Neumann function of order 2 and $\gamma_n$ is an arbitrary constant. The KK graviton modes (that is, $m_n$) and $\gamma_n$ can be found from the continuity condition of the wavefunction at the two orbifold fixed points, that is, at $u=0$ and $u=b_u$. The continuity condition at $u=0$ implies $\gamma_n \ll 1$ as $m_n/K \ll 1$. This leads to
\begin{equation}\label{4.26}
    \Psi_n(u) \simeq N_ne^{2A(u)}J_2\left(\frac{m_n}{K}e^{A(u)}\right)
\end{equation}
The continuity condition at $u=b_u$ provides the following
\begin{equation}\label{4.27}
    J_1(x_n) = 0\;\;\;\;\;\text{with}\;\;\;\;\;x_n = \frac{m_n}{K}e^{A(b_u)}
\end{equation}
where, $b_u$ is defined as
\begin{equation}\label{4.28}
    b_u \equiv \int_0^b dy\;B(y) = b + \frac{1}{2}\varepsilon b^2
\end{equation}
From the above relations, we can easily read off the KK mass spectrum of massive gravitons
\begin{equation}\label{4.29}
    m_n \simeq Kx_ne^{-A(y=b)}
\end{equation}
The normalization of the graviton wavefunctions $\Psi_n$'s can be obtained from the orthonormality condition stated in eq.(\ref{4.4})
\begin{equation}\label{4.30}
    N_n =\frac{\sqrt{2K}\;e^{-Kb_u}}{J_2(x_n)}
\end{equation}
Finally in $y$ coordinates, we can express approximate analytic closed form solutions of the massive graviton wavefunctions as follows
\begin{equation}\label{4.31}
    \Psi_n(y) \simeq \frac{\sqrt{2K}\;e^{-A(y = b)}}{J_2(x_n)}e^{2A(y)}J_2\left(\frac{m_n}{K}e^{A(y)}\right)\;\;\;\;\;;\;\;n=1,2,3,...
\end{equation}

\noindent Though we have already obtained the useful expressions of zero mode and massive mode wavefunctions along with the graviton mass spectrum, it is instructive to do the same analysis by an alternative method - quantum mechanical perturbation theory. Although it is not possible to obtain a closed form graviton wavefunctions by this method, but we will get the same spectrum by this method as well.

\subsection{Analysis II: QM-like Perturbation Theory}\label{analysis2}

 In order to solve the first order perturbation problem, we cast the full problem in a familiar setting of 1D quantum mechanical problem. Therefore, we rewrite the eq.(\ref{4.7}) as follows
\begin{equation}\label{4.32}
(\mathcal H_0 + \varepsilon\mathcal H_1)\ket{\zeta_n} = m_n^2(w_0 + \varepsilon w_1)\ket{\zeta_n} \;\;\;\;\;;\;\; n = 0,1,2,3,...
\end{equation}
with
\begin{equation}\label{4.33}
\mathcal H_0 \equiv -\partial_y^2 + 4K^2 \;\;;\;\; \mathcal H_1 \equiv F_1(y)\;\;;\;\;w_0 \equiv e^{2Ky} \;\;;\;\; w_1 \equiv e^{2Ky}F_2(y)
\end{equation}
where,
\begin{equation}\label{4.34}
F_1(y) = 8Kky + (2K-2k) \;\;\;\;;\;\;\; F_2(y) = 2y + ky^2
\end{equation}
Note that we adopted the Dirac's bra-ket notation for the ease of further calculations. In the bra-ket notation inner product is defined as follows
\begin{equation}\label{4.35}
    \braket{\zeta_n|\zeta_m} \equiv \int_0^\infty dy\;\zeta_n(y)\zeta_m(y)
\end{equation}
One can easily verify that the unperturbed states $\ket{\zeta^{(0)}_n}$ in eq.(\ref{4.12}) are normalized, that is, $\bra{\zeta^{(0)}_n}w_0\ket{\zeta^{(0)}_n} = 1$. Now expanding $\ket{\zeta_n}$ and $m^2_{n}$ in the perturbation series up to first order in our perturbative parameter $\varepsilon$ 
\begin{align}\label{4.36}
\ket{\zeta_n} &= \ket{\zeta^{(0)}_n} + \varepsilon\ket{\zeta^{(1)}_n} + \mathcal O(\varepsilon^2) \nonumber \\
m^2_n &= m^{2(0)}_n + \varepsilon m^{2(1)}_n + \mathcal O(\varepsilon^2)
\end{align}
we get the first order correction $m_{(n)}^{2(1)}$ and $\ket{\zeta_n^{(1)}}$ as follows
\begin{align}\label{4.37}
m_n^{2(1)} &= \frac{\bra{\zeta^{(0)}_n}\mathcal H_1\ket{\zeta^{(0)}_n} - m^{2(0)}_n\bra{\zeta^{(0)}_n}w_1\ket{\zeta^{(0)}_n}}{\bra{\zeta^{(0)}_n}w_0\ket{\zeta^{(0)}_n}} \\
\ket{\zeta_n^{(1)}} &= \sum_{m\neq n} \frac{\bra{\zeta_m^{(0)}}\mathcal H_1\ket{\zeta_n^{(0)}} - m_n^{2(0)}\bra{\zeta_m^{(0)}}w_1\ket{\zeta_n^{(0)}}}{m_n^{2(0)} - m_m^{2(0)}} \ket{\zeta_m^{(0)}} \label{4.38}
\end{align}
Note that quantities with superscript $^{(1)}$ denotes the first correction to their corresponding unperturbed part. In the subsequent subsections, we are going to explore the consequences of the higher curvature effects in the unperturbed KK modes through their modifications in the graviton spectrum as well as the graviton wavefunctions at the leading order.

\subsubsection{Correction to Unperturbed Zero Mode}\label{unp_zero}

For unperturbed zero graviton mode, $m^{2(0)}_0 = 0$. Therefore the first order correction in mass square can be simply given as from eq.(\ref{4.37}) assuming $Kb \gg 1$
\begin{equation}\label{4.39}
m_0^{2(1)} = \bra{\zeta^{(0)}_0}\mathcal H_1\ket{\zeta^{(0)}_0} = 2K\int_0^\infty dy\;e^{-4Ky}\left[8Kky + (2K-2k)\right] = K
\end{equation}
Therefore, the unperturbed zero graviton mode picks up a mass correction due to the higher curvature effect. And up to the first order perturbation, this is given by
\begin{equation}\label{4.40}
m^2_0 = \varepsilon K
\end{equation}
and similarly using eq.(\ref{4.38}), we can get the correction to the unperturbed zero mode graviton wavefunction as follows
\begin{equation}\label{4.41}
|\zeta_0^{(1)}\rangle = -e^{-Kb} \sum_{m=1}^{\infty}
\frac{1}{J_2(x_m)} \left[ k b^2 + P_m b + Q_m \right]
|\zeta_m^{(0)}\rangle
\end{equation}
where, the coefficients $P_m$ and $Q_m$ are defined in Appendix \ref{B}. \\ 

\noindent It is important to emphasize that eq.~(\ref{4.40}) should not be interpreted as implying that the physical massless graviton of the higher-curvature theory acquires a non-vanishing mass. In perturbation theory, the quantity $m_0^{2(1)}$ represents the correction to the eigenvalue associated with the unperturbed (shifted) RS zero mode state. However, the physical spectrum of the full higher curvature theory must be determined from the complete eigenvalue problem rather than by tracking the evolution of a particular unperturbed state. Indeed, as shown in the previous subsection, the full theory admits a normalizable massless graviton mode. Therefore, the physical zero mode of the higher-curvature theory need not coincide with the unperturbed (shifted) RS zero mode, and the result in eq.~(\ref{4.40}) should be understood only as a perturbative shift of the unperturbed basis state rather than as a mass generation mechanism for the physical graviton.

\subsubsection{Correction to Unperturbed Massive Modes}\label{unp_massive}

Using eq.(\ref{4.37}), we can calculate the first order correction in the massive unperturbed graviton modes as follows
\begin{equation}\label{4.42}
    m_n^{2(1)} = \bra{\zeta^{(0)}_n}\mathcal H_1\ket{\zeta^{(0)}_n} - m^{2(0)}_n\bra{\zeta^{(0)}_n}w_1\ket{\zeta^{(0)}_n}
\end{equation}
where, we know the masses of the unperturbed massive modes are $m_{n}^{(0)} = Kx_ne^{-Kb}$ (for $\varepsilon = 0$). The masses of the massive unperturbed KK modes up to a leading order perturbation (assuming $kb \gg 1$) is the following
\begin{equation}\label{4.43}
    m_n^2 \simeq K^2x_n^2e^{-2Kb}\left[1 - \varepsilon kb^2\right]
\end{equation}
Note that the above spectrum can be identified with the one derived earlier in eq.(\ref{4.29}) for small $\varepsilon$ and we will get $m_n \simeq Kx_ne^{-A(y=b)}$. Similar to the zero mode correction, using eq.(\ref{4.38}), we can get the correction to the unperturbed massive graviton wavefunctions as follows
\begin{equation}\label{4.44}
|\zeta_n^{(1)}\rangle = 
C_{n0}\ket{\zeta_0^{(0)}}
+ \sum_{m\neq 0,n} ( \mathcal P_{nm}\, b^2 + \mathcal Q_{nm}\, b + \mathcal L_{nm})\ket{\zeta_m^{(0)}}
\end{equation}
where, the coefficients $C_{n0}$ and $\mathcal P_{nm},\mathcal Q_{nm},\mathcal L_{nm}$ are defined in the Appendix \ref{B}. Note that it is not possible to bring the wavefunctions in eq.(\ref{4.41}) and eq.(\ref{4.44}) in a closed form to work with. Therefore for the discussions in following subsections, we restrict ourselves to the approximate wavefunctions obtained in eq.(\ref{4.17}) and (\ref{4.31}).

\subsection{Coupling of KK Gravitons with Standard Model Fields on the TeV Brane}\label{coupling}

We consider the interaction between the KK graviton modes and the standard model (SM) matter fields residing on the TeV brane, located at $y = b$. The solution for tensor fluctuations that appear at the TeV brane can be obtained by substituting the solutions for $\Psi_n(y)$ for $n=0,1,2,...$ at $y = b$ (see eq.(\ref{4.17}) and (\ref{4.31}))
\begin{equation}\label{4.45}
    h_{\mu\nu}(x,y = b) = \sum_{n=0}^\infty \hat h_{\mu\nu}^{(n)}(x)\Psi_n(y = b) = \frac{M_*^{3/2}}{M_{Pl}}\hat h^{(0)}_{\mu\nu} + \sqrt{2K}\;e^{A(b)}\sum_{n=1}^\infty \hat h_{\mu\nu}^{(n)}(x)
\end{equation}
The interaction Lagrangian (of gravitons and SM fields) in the 4D effective theory can be expressed as
\begin{equation}\label{4.46}
    \mathcal L_{TeV} = \frac{1}{M_*^{3/2}}T^{\mu\nu}(x)h_{\mu\nu}(x,y=b)
\end{equation}
where, $T^{\mu\nu}(x)$ is the energy-momentum tensor of the SM matter fields on the TeV brane. Now we use the relation between 5D Planck scale ($M_*$) and 4D Planck scale ($M_{Pl}$) as shown in eq.(\ref{2.20}) and this leads to
\begin{equation}\label{4.47}
    \mathcal L_{TeV} = \frac{1}{M_{Pl}}T^{\mu\nu}(x)\hat h_{\mu\nu}^{(0)} + \frac{1}{M_{Pl}\;e^{-A(b)}}\beta\sum_{n=1}^\infty T^{\mu\nu}(x)\hat h_{\mu\nu}^{(n)}
\end{equation}
where, $\beta = \beta(\alpha,b_0)$ is a parameter appearing in the coupling which depends on our free parameters $\alpha, b_0$ and it is given by
\begin{align}\label{4.48}
\beta(\alpha,b_0) = \sqrt{2K}\Bigg[&\frac{1}{2k}\left(1 - e^{-2Kb - \varepsilon kb^2}\right) \nonumber \\
&+ \frac{1}{2}\sqrt{\frac{\pi}{\varepsilon k}}\left(1 - \frac{K}{k}\right)e^{K^2/\varepsilon k}\Big\{\mathbf{erf}\left(\sqrt{\varepsilon k}\;b + \frac{K}{\sqrt{\varepsilon k}}\right) - \mathbf{erf}\left(\frac{K}{\sqrt{\varepsilon k}}\right)\Big\}\Bigg]^{1/2}
\end{align}
This parameter $\beta(\alpha,b_0)$ gives the modification of the coupling of the KK graviton with the SM matter fields from that evaluated in the original 5D RS model.

\subsection{Phenomenological Implications}\label{pheno}

In the previous subsection, we discussed about the coupling of KK graviton modes with the SM matter fields confined to TeV brane in the context of higher curvature back-reacted model. We have also introduced the $\beta$ parameter which parametrizes the modifications in the coupling of KK gravitons with SM matter fields. \\

\begin{figure}[h!]
    \centering
    
    \begin{subfigure}[b]{0.45\textwidth}
        \centering
        \includegraphics[width=\textwidth]{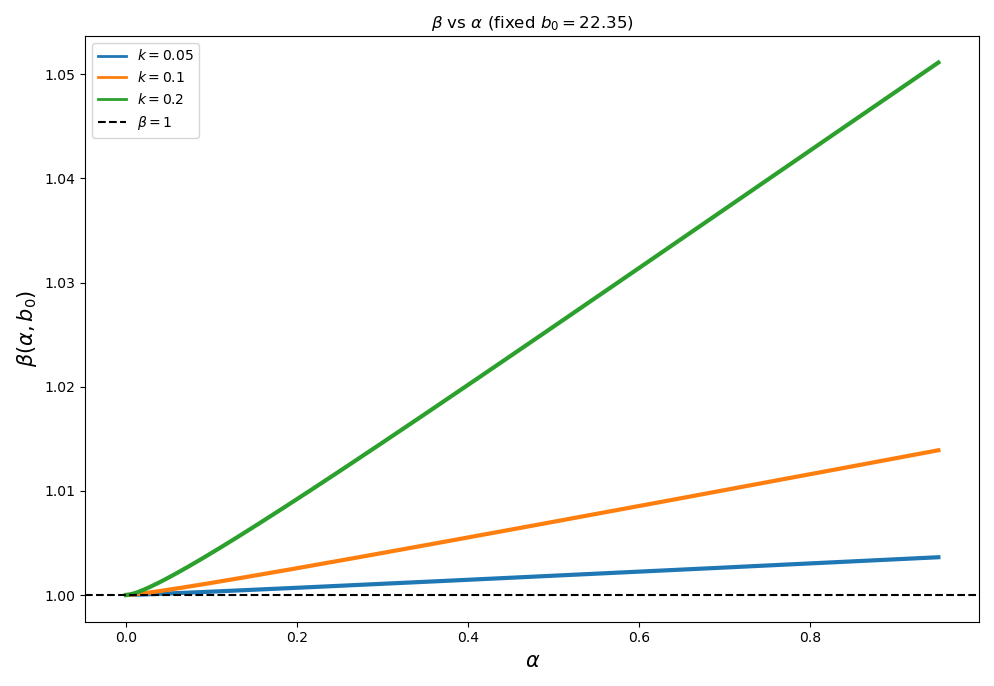}
        \caption{}
        \label{Fig1a}
    \end{subfigure}
    \hspace{1cm}
    \begin{subfigure}[b]{0.45\textwidth}
        \centering
        \includegraphics[width=\textwidth]{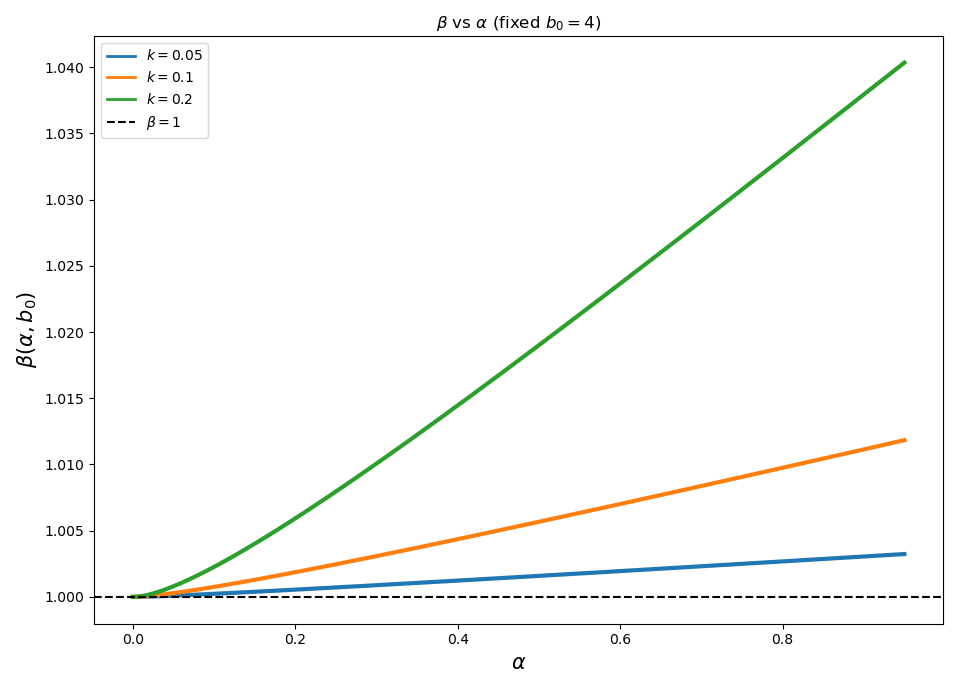}
        \caption{}
        \label{Fig1b}
    \end{subfigure}
    
    \caption{$\beta(\alpha,b_0)$ vs $\alpha$ plots for fixed values of $b_0$. Chosen values: $b_0 = 22.35$ for plot in (a) and $b_0 = 4$ for plot in (b). The blue, orange and green curves corresponds to $k=0.05,0.1,0.2$ respectively (in Planck units).}
    \label{Fig1}
\end{figure}

\begin{figure}[h!]
    \centering
    
    \begin{subfigure}[b]{0.45\textwidth}
        \centering
        \includegraphics[width=\textwidth]{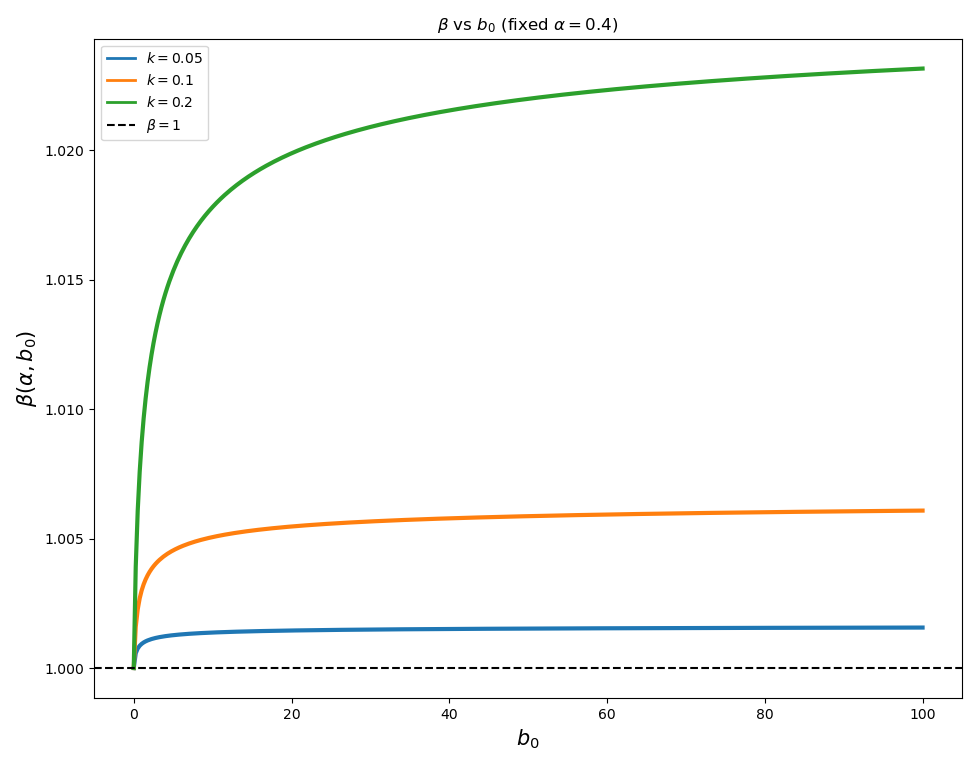}
        \caption{}
        \label{Fig2a}
    \end{subfigure}
    \hspace{1cm}
    \begin{subfigure}[b]{0.45\textwidth}
        \centering
        \includegraphics[width=\textwidth]{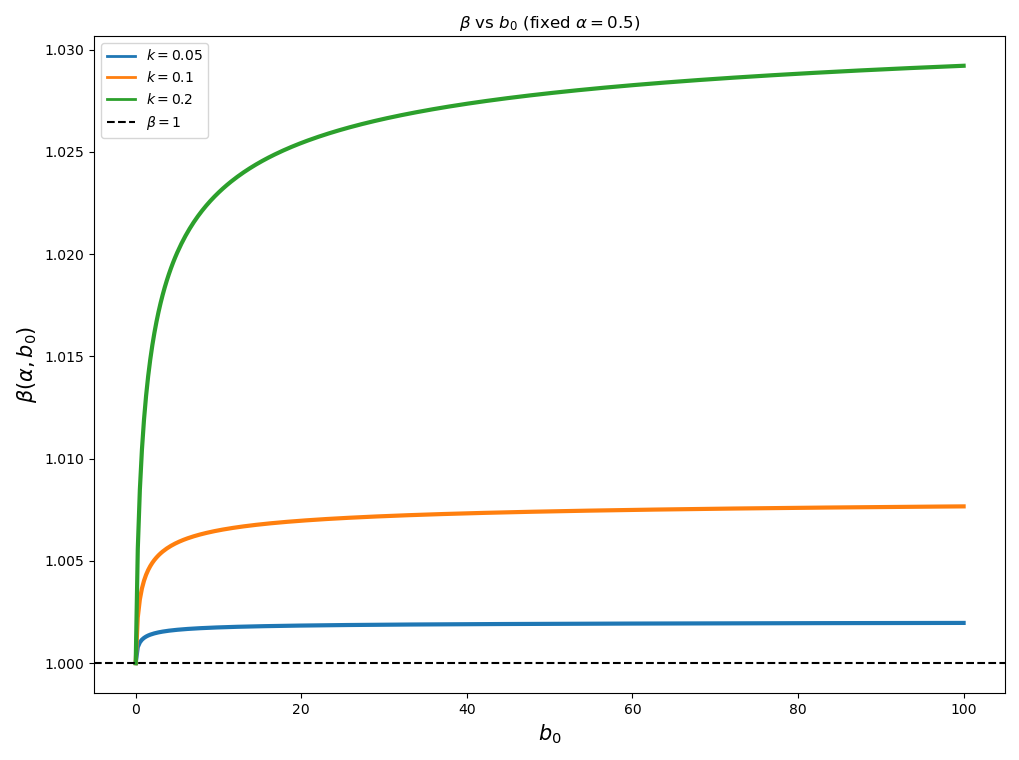}
        \caption{}
        \label{Fig2b}
    \end{subfigure}
    
    \caption{$\beta(\alpha,b_0)$ vs $b_0$ plots for fixed values of $\alpha$. Chosen values: $\alpha = 0.4$ for plot in (a) and $\alpha = 0.5$ for plot in (b). The blue, orange and green curves corresponds to $k=0.05,0.1,0.2$ respectively (in Planck units).}
    \label{Fig2}
\end{figure}

\begin{figure}[h!]
    \centering
    
    \begin{subfigure}[b]{0.45\textwidth}
        \centering
        \includegraphics[width=\textwidth]{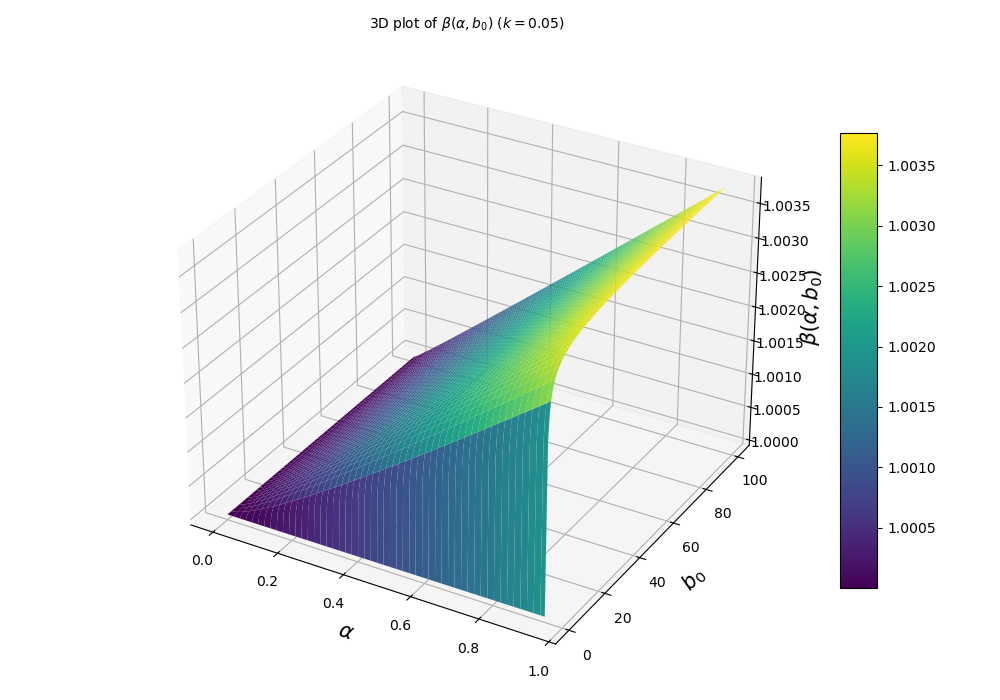}
        \caption{}
    \end{subfigure}
    \hspace{1cm}
    \begin{subfigure}[b]{0.45\textwidth}
        \centering
        \includegraphics[width=\textwidth]{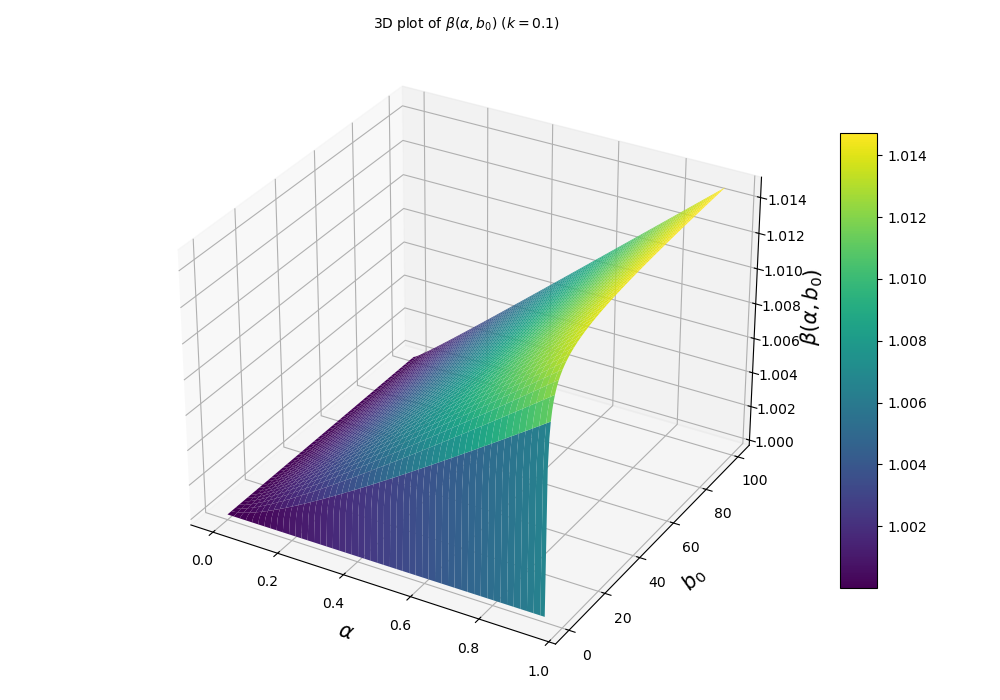}
        \caption{}
    \end{subfigure}
    
    \caption{3D plots of $\beta(\alpha,b_0)$ vs $(\alpha,b_0)$. Chosen value of $k$: $k = 0.05$ for plot in (a) and $k = 0.1$ for plot in (b) (in Planck units).}
    \label{Fig3}
\end{figure}

\begin{figure}[h!]
    \centering
    
    \begin{subfigure}[b]{0.45\textwidth}
        \centering
        \includegraphics[width=\textwidth]{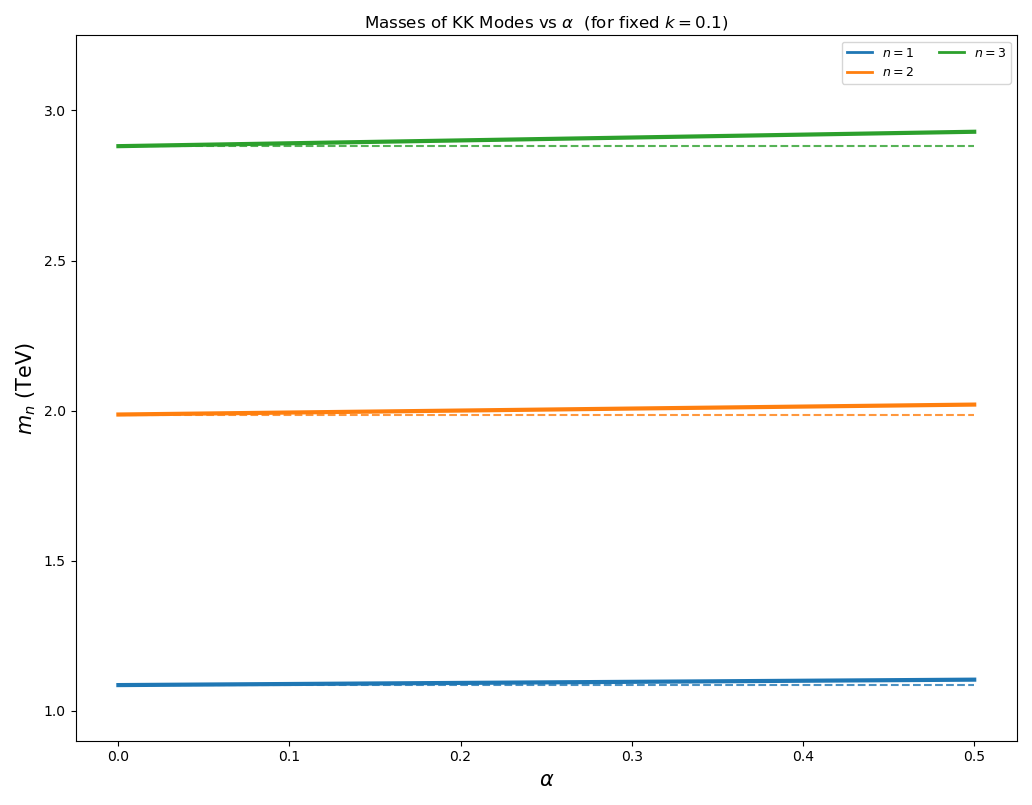}
        \caption{}
    \end{subfigure}
    \hspace{1cm}
    \begin{subfigure}[b]{0.45\textwidth}
        \centering
        \includegraphics[width=\textwidth]{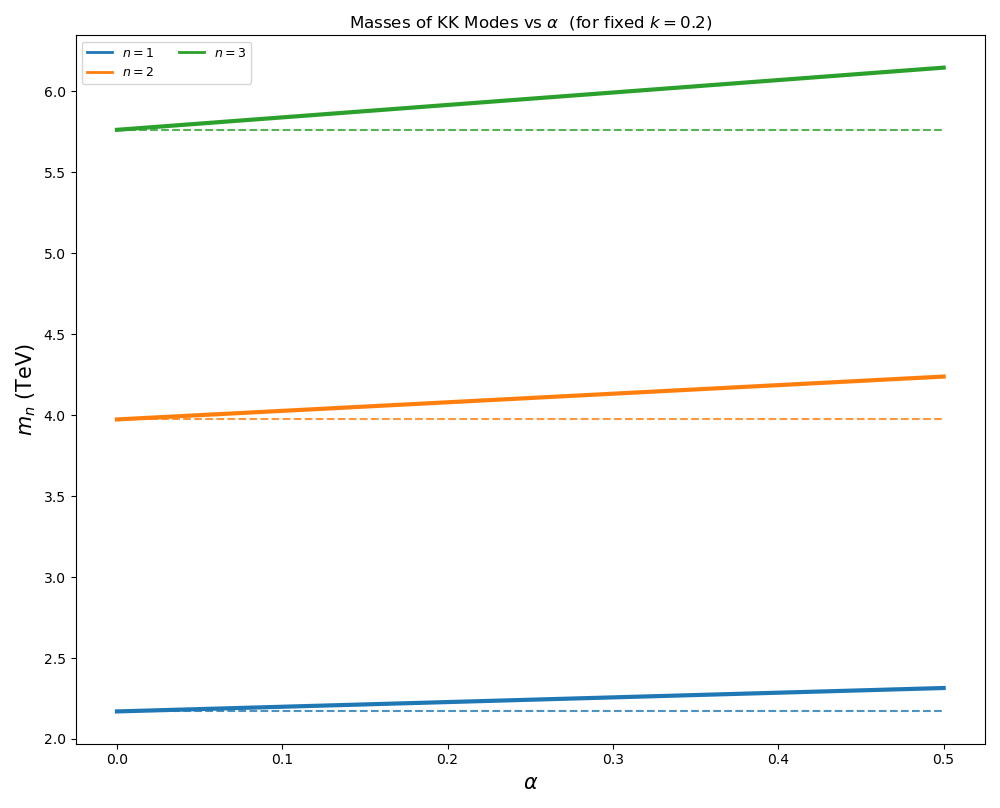}
        \caption{}
    \end{subfigure}
    
    \caption{Plots of masses of first three KK graviton modes with the higher curvature parameter $\alpha$. Chosen values of $k$: $k = 0.1$ for the plot in (a) and $k = 0.2$ for the plot in (b) (in Planck units). The blue, orange and green curves corresponds to $n=1,2,3$ modes, respectively. And the dashed lines are the corresponding RS masses ($\alpha = 0$).}
    \label{Fig4}
\end{figure}

\begin{figure}[h!]
    \centering
    \includegraphics[width=0.55\linewidth]{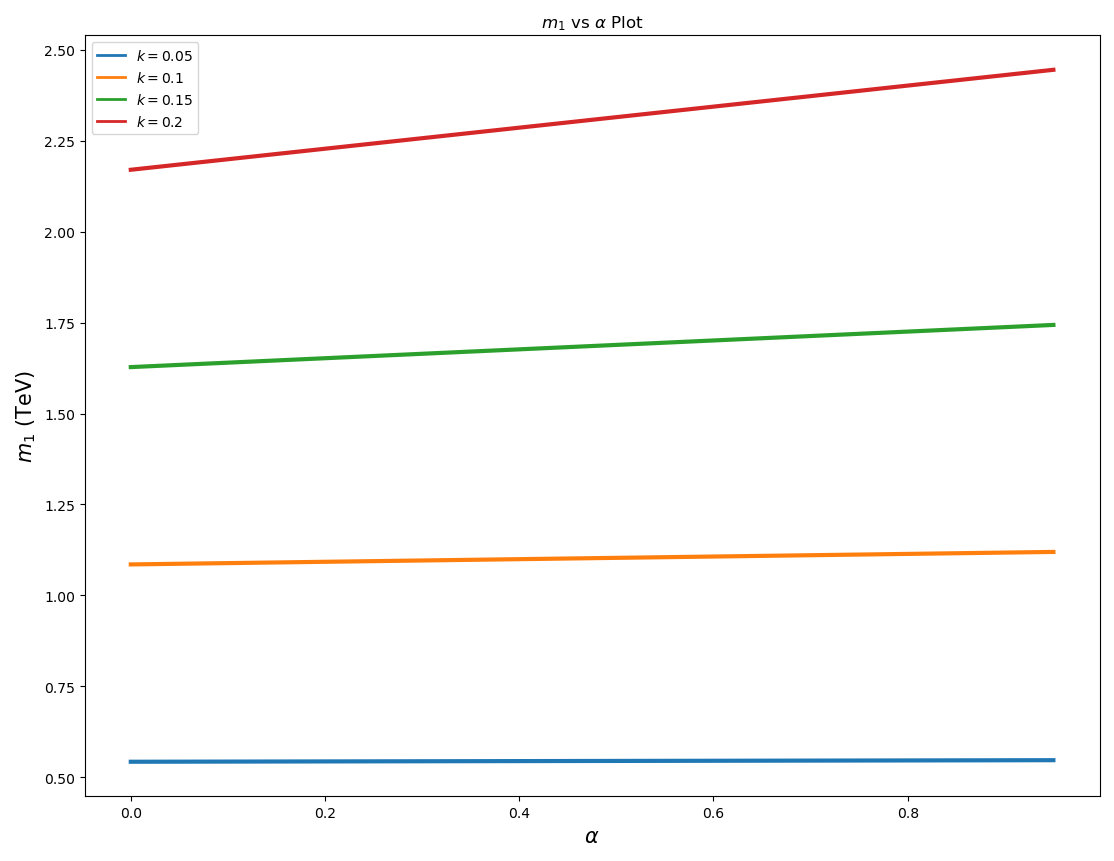}
    \caption{Plot of lightest KK mode mass vs $\alpha$. The blue, orange, green, red and violet curves corresponds to $k = 0.05,0.10,0.15,0.20$ respectively (in Planck units).}
    \label{Fig5}
\end{figure}

\noindent In order to address the gauge hierarchy problem, we assume that the modified warp factor produces the same warping as in the original RS scenario. Hence
\begin{equation}\label{4.49}
    A(y=b) - A(y=0) = 36
\end{equation}
The above condition produces the following correlation between the parameters $\alpha,b_0,k$ and $b$
\begin{equation}\label{4.50}
    kb + \frac{10}{3}\alpha k^3b + \frac{1}{2}\alpha b_0 kb^2 = 36
\end{equation}
The above equation can be solved for a given parameter set $(k,\alpha,b_0)$ to get the brane separation $b$. Following this exact procedure, we plot $\beta$ as a function of the free parameters $\alpha$ and $b_0$ for some fixed values of $k$ (see Figures \ref{Fig1}, \ref{Fig2} and \ref{Fig3}). As we are fixing the warping at TeV brane to be the same as standard RS scenario (that is, $A(b) = 36$), then the spectrum of massive gravitons $m_n = Kx_ne^{-A(b)}$ does not depend on the back-reaction parameter $b_0$. It only depends on the higher curvature parameter $\alpha$ through $K = k + \frac{10}{3}\alpha k^3$. Below we present plots of masses of different massive modes with the higher curvature parameter $\alpha$ (see Figure \ref{Fig4}). Also we plot only the mass of lightest massive KK graviton mode ($n=1$) with the higher curvature parameter $\alpha$ for different $k$ values in a separate figure (see Figure \ref{Fig5}). \\


\noindent We know $\beta(\alpha,b_0) = 1$ is the standard RS value, means no modification in coupling of KK gravitons with SM matter fields at the TeV brane. That's why we put a $\beta = 1$ lines in Figure \ref{Fig1} and \ref{Fig2}, so that the modification is clearly visible. It is clear from these two figures that $\beta$ is of increasing nature, it never becomes less than $1$. From Figure \ref{Fig1}, we can see $\beta$ is almost increasing linearly with increasing $\alpha$ for a fixed $b_0$ value. Also note that increasing $k$ values increases the slope of the plots, in other words, $\beta$ increases more for larger $k$ values. And comparing Figure \ref{Fig1a} and \ref{Fig1b}, we can see that for large fixed $b_0$ value, $\beta$ increases slightly more than the smaller fixed $b_0$ value. It is better to take an approximate value of $\beta$ and see what it conveys. From Figure \ref{Fig1a}, for $k=0.2$ and $\alpha = 0.4$, $\beta \simeq 1.02$. That means the coupling gets enhanced by $\approx 2\%$ for this case. Now from Figure \ref{Fig2}, we can see with increasing $b_0$ value (for fixed $\alpha$), $\beta$ at first increases sharply, but soon it saturates to a fixed value for large $b_0$ values. And with the increasing $k$ values, $\beta$ is increasing. From Figure \ref{Fig2a}, the $k=0.1$ curve saturates to a value $\beta\simeq1.005$, which indicates a tiny enhancement $\approx 0.5\%$ in coupling of KK gravitons with SM fields. \\

\noindent Figure \ref{Fig4} indicates that the masses of different KK graviton modes increases linearly with $\alpha$ and the large values of $k$ increases the slopes of the curves. As we have demanded the same warping (that is, $A(b) = 36$), the masses are independent of the back-reaction parameter $b_0$. Figure \ref{Fig5} shows the variation of $m_1$ (mass of the lightest KK mode) with respect to $\alpha$ for different $k$ values. Note that for larger values of $k$, there is an significant lifting of mass of the lightest KK graviton mode. For $k=0.2$, $m_1^{RS} \simeq 2.17$TeV and $m_1^{mod} \simeq 2.31$TeV, which indicates approximately $\approx 6.5\%$ enhancement in the mass of the lightest KK graviton mode. There is an interesting point to note: the coupling gets enhanced and as well as $m_1$ gets enhanced. In this scenario $\beta^2/m_1^2$ is an useful quantity to evaluate and compare it with the standard RS value in order to extract something meaningful. Below we provide a table containing the (dimensionless) ratio $\beta^2/m_1^2$ for different cases.

\begin{table}[h!]
    \centering
\begin{tabular}{c||c|c|c|c|c}
$k$ & $m_1^{RS}$ in TeV & $m_1^{mod}$ in TeV & $\beta_{mod}$ & $(\beta^2/m^2_{1})_{RS} (\times 10^{32})$ & $(\beta^2/m_1^2)_{mod}(\times 10^{32})$ \\ \hline\hline
$0.05$ & 0.54255 & 0.54481 & 1.00175 & 5.06382 & 5.03947\\       
$0.10$ & 1.08510 & 1.10319 & 1.00649 & 1.26596 & 1.24075\\
$0.20$ & 2.17021 & 2.31489 & 1.02301 & 0.31649 & 0.29111\\
\end{tabular}
    \caption{The table shows some typical values for $m_1,\beta$ and $\beta^2/m_1^2$ values for both RS and modified (mod) cases for different values of $k$ (in Planck units). Here parameter values $\alpha = 0.4$ and $b_0 = 10$ are taken.}
    \label{tab:placeholder}
\end{table}
\noindent From the above table, see the case of $k=0.2$, we get
\begin{equation}\label{4.51}
    \frac{(\beta^2/m_1^2)_{mod}}{(\beta^2/m_1^2)_{RS}} \approx 0.92 \implies \approx8\%\;\text{suppression in }\beta^2/m_1^2 
\end{equation}
Before progressing further, we now want to discuss about the detection of gravitons in colliders. Basically, there are two methods by which gravitons can affect the collider physics. First one is an \textit{indirect effect}, in which the scattering amplitudes of SM particles are modified through virtual graviton exchange. Such contributions can modify the production cross-sections of various SM processes. Another one is a \textit{direct effect}, in which an on-shell graviton decays into SM particles, such as dilepton or diphoton final states. Such processes are characterized by the decay widths of the KK gravitons into various SM final states. Searches for these resonant decay channels have been extensively performed by the ATLAS and CMS collaborations in LHC. The invariant amplitude for virtual graviton exchange processes is roughly proportional to $\sim \beta^2/m_n^2$ and the cross-section goes as $\sim\beta^4/m_n^4$, whereas the partial decay width of a KK graviton channel scales approximately as $\sim\beta^2m_n^3$. Moreover, the branching ratio (BR) of a KK graviton decay channel is determined by the ratio of its partial decay width to the total decay width.\\

\noindent Therefore, we can see that in the back-reacted higher curvature scenario, mass of the lightest KK graviton mode increases more significantly than its coupling, leading to a modest suppression in $\beta^2/m_1^2$. This suggests a significant reduction in collider cross-sections for processes involving virtual graviton exchange. On the other hand, the simultaneous enhancement of $\beta$ and the lightest KK graviton mass $m_1$ relative to the original RS model leads to a significant increase in the dilepton and diphoton decay widths. Furthermore, since higher curvature effects enhance the KK graviton masses and total decay widths, additional decay channels may become kinematically accessible, potentially modifying the branching ratios relative to the original RS model. Overall, our model predicts observable modifications to both the production and decay signatures of the lightest KK graviton relative to the original RS scenario. A quantitative assessment of these effects requires a dedicated collider analysis, which we leave for future work.\\

\noindent Below we provide a brief qualitative discussion about the bulk geometry and how it affects the phenomenology of the KK gravitons at the TeV brane.

\subsubsection*{Geometric Origin of the Collider Modifications}

\begin{figure}[h!]
    \centering
    \includegraphics[width=0.65\linewidth]{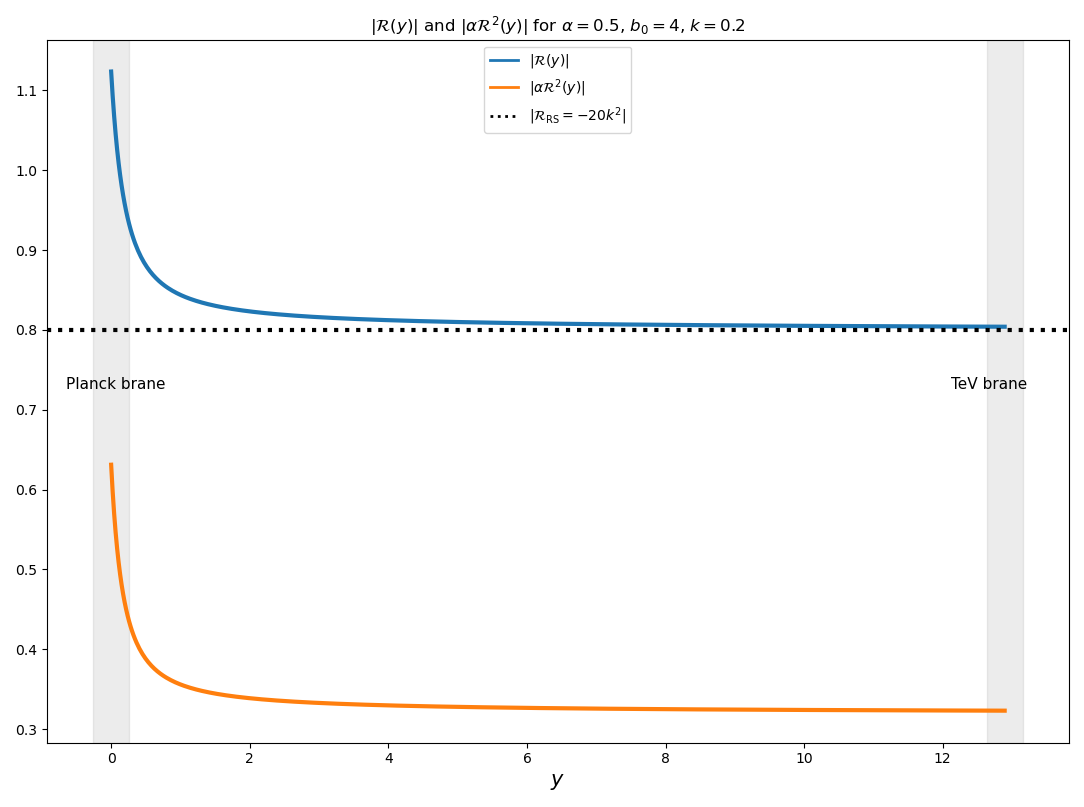}
    \caption{Plots of $|\mathcal{R}(y)|$ and $|\alpha\mathcal R^2(y)|$ as a function of $y$. The dotted line shows the original RS value $|\mathcal R = -20k^2|$. We have taken $\alpha = 0.5$, $b_0 = 4$ and $k = 0.2$.}
    \label{Fig6}
\end{figure}
\noindent We can explain the above results from a different viewpoint. Recall the 5D curvature scalar in eq.(\ref{2.9}). Using the expressions in eq.(\ref{2.15}), we can express the curvature scalar as a function of $y$. In Figure \ref{Fig6}, we present a plot of the curvature scalar $|\mathcal{R}(y)|$ and $|\alpha\mathcal R^2(y)|$ in order to see the higher curvature effects throughout the bulk for a fixed parameter choice. We can see that the curvature scalar significantly deviates from the original RS value near the Planck brane ($y=0$). But soon it comes close to the original RS value near the TeV brane ($y=b$). We have the numbers to compare (for the parameter choice used in Figure \ref{Fig6}): $|\mathcal R(b)| \simeq 0.804$ and $|\mathcal R(b)|_{RS} = 0.800$, which differs by only $\approx 0.5\%$ as compared to the RS situation. And at $y=b$, $\alpha\mathcal R^2/|\mathcal R| \simeq 0.402$, which implies $\alpha\mathcal R^2$ is $\approx40\%$ of $|\mathcal R|$ at the TeV brane, which is a comparable higher curvature effect at the TeV brane. So the concluding interpretation is the following: The geometrical effects are prominent near the Planck brane. The curvature rapidly approaches the original RS value towards the TeV brane, indicating that the warped geometry remains remarkably close to the original RS $AdS_5$ geometry near the TeV brane. That's why the KK graviton spectrum and couplings with SM fields at TeV brane receive only modest higher curvature back-reacted corrections inspite of having a comparable higher curvature term present in the action. Nevertheless, these corrections can still lead to appreciable modifications in collider observables, such as production cross-sections and decay widths, owing to their nontrivial dependence on the KK graviton masses and couplings.\\

\noindent Below we provide a justification and validity of our chosen parameter space used in the phenomenology of KK gravitons from a naive dimensional analysis point of view and show that the effective field theory is well defined for our case. We will also discuss the probable modification to our result for inclusion of more higher-order curvature operators.

\subsection{Higher-Order Curvature Operators and Convergence of EFT}\label{EFT}

\begin{figure}[h!]
    \centering
    \includegraphics[width=0.75\linewidth]{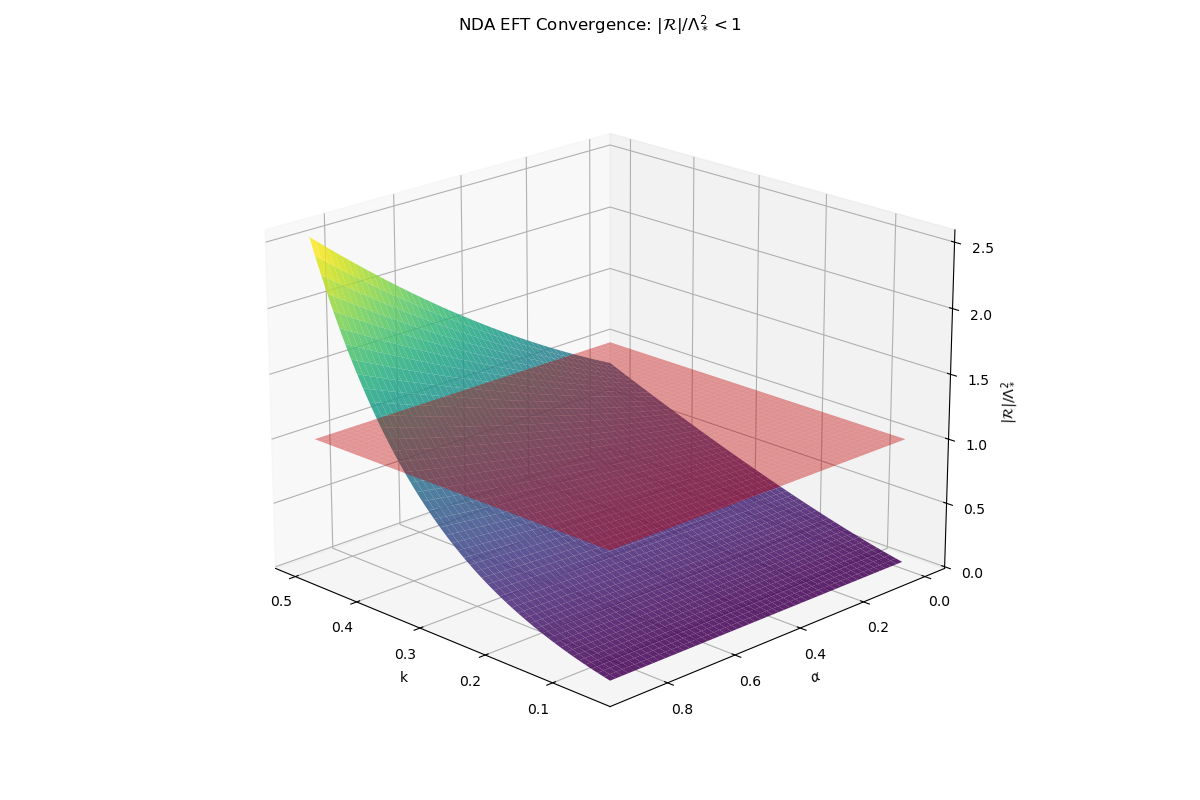}
    \caption{3D plot of $|\mathcal R/\Lambda_*^2|$ vs $(k,\alpha)$ in Planck units. The red plane is $|\mathcal R/\Lambda_*^2| = 1$ plane which intersects the 3D plot. The allowed $(k,\alpha)$ values are constrained by $|\mathcal R/\Lambda_*^2| < 1$, that is, below the red plane.}
    \label{fig:EFT_NDA}
\end{figure}

The 5D $f(\mathcal R)$ gravity is a non-renormalizable theory, hence is an effective field theory (EFT) involving higher-order curvature operators. This EFT has a certain cut-off scale, denoted as $\Lambda_*$ (mass dimension is 1). At the energy scale $\Lambda_*$, the EFT breaks down and the 5D gravity theory becomes strongly coupled. The higher curvature operators in $f(\mathcal R)$ gravity action are suppressed by powers of $\mathcal R/\Lambda_*^2$, that is
\begin{equation}\label{4.52}
    S = M_*^3\int d^5x \sqrt{-g_5}\;\left[-\Lambda + \mathcal R + \frac{c_2}{\Lambda_*^2}\mathcal R^2 + \frac{c_3}{\Lambda_*^4}\mathcal R^3 + \cdot\cdot\cdot\right]
\end{equation}
where, $c_2,c_3,...$ are $\mathcal O(1)$ dimensionless numbers. The naive dimensional analysis (NDA) estimate is given by~\cite{N1,N2}
\begin{equation}
    \frac{\Lambda_*^3}{24\pi^3} \sim M_*^3
\end{equation}
We can show that loop effects and local higher-dimensional operators in the 5D theory are also suppressed by a similar factor. In Planck units, we can obtain $\Lambda_* \sim (3\pi^2k)^{1/3}$ and the Ricci scalar for our metric in eq.(\ref{2.5}) is
\begin{equation}
    \mathcal R = -20\left[k + \frac{10}{3}\alpha k^3\right]^2 + \mathcal O(\varepsilon)
\end{equation}
Ignoring the $\mathcal O(\varepsilon)$ contribution in Ricci scalar, we can obtain the allowed parameter space of $(k,\alpha)$. In Figure \ref{fig:EFT_NDA}, we have plotted a 3D plot of $|\mathcal R/\Lambda_*^2|$ vs $(k,\alpha)$, which gives the allowed parameter space by the requirement $|\mathcal R/\Lambda_*^2| < 1$ for the convergence of the EFT. Thus, all our chosen values of $k$ and $\alpha$ for the phenomenological discussion of KK gravitons are perfectly fine and EFT will converge for each of the cases. \\

\noindent All our calculations and results are so far considering $f(\mathcal R) = \mathcal R + \alpha\mathcal R^2 - \Lambda$. But we can in principle choose more higher-order curvature operators, like $\mathcal R^3,\mathcal R^4,...$ terms. We will now briefly discuss what will going to happen for that scenario. Suppose in addition we have a $\beta\mathcal R^3$ term present in the $f(\mathcal R)$. Then from the action in eq.(\ref{4.52}), we can identify $\alpha$ and $\beta$ as
\begin{equation}
    \alpha \equiv \frac{c_2}{\Lambda_*^2}\;\;\;;\;\;\; \beta \equiv \frac{c_3}{\Lambda_*^4}
\end{equation}
As $c_2,c_3 \sim \mathcal O(1)$, we can further have
\begin{equation}
    \beta \sim \alpha^2
\end{equation}
From our previous analysis in section \ref{modified}, we essentially present the corrections or modifications in KK graviton masses and couplings with SM fields at $\mathcal O(\alpha)$ (in Planck units) for $f(\mathcal R) = \mathcal R + \alpha\mathcal R^2 - \Lambda$ gravity. Incorporating the $\beta\mathcal{R}^3$ term on top of our obtained results as a perturbative background will yield $\mathcal{O}(\alpha^2)$ corrections. Similarly, the general inclusion of an $\mathcal{R}^n$ operator will introduce $\mathcal{O}(\alpha^{n-1})$ corrections. Because these higher-order corrections are strictly subdominant from naive dimensional analysis, our results obtained at $\mathcal{O}(\alpha)$ represent a physically valid and well-controlled perturbative expansion, ensuring that the EFT remains well-defined and its series converges properly within our chosen parameter space.

\section{Discussion}\label{discussion}

We have perturbatively found the solutions of warp factors (that is, $A(y)$ and $B(y)$) by solving the modified higher curvature field equations for the model $f(\mathcal R) = \mathcal R + \alpha\mathcal R^2 - \Lambda$. Further we showed the exact relation between the 4D Planck scale $M_{Pl}$ and the 5D Planck scale $M_*$ for our chosen higher curvature scenario. We briefly reviewed the well-studied Randall-Sundrum Kaluza-Klein graviton problem and solved it explicitly using a novel method of Euclidean path integrals. We have  obtained analytic closed form solutions of the graviton wavefunctions as well as the KK graviton spectrum. However, we have obtained the modified KK graviton spectrum which  depends on the higher curvature parameter $\alpha$. The mass of the KK graviton modes are shown to increase linearly with the higher curvature parameter. We have obtained a modification to the coupling of KK graviton modes with the SM matter fields which is dependent on the free parameters of our model $\alpha,b_0$ ($b_0$ is the back-reaction parameter). We showed the variation of this modification of the coupling with respect to $\alpha$ and $b_0$ for various choices of $AdS_5$ curvature scale $k$ and these effectively indicate an enhancement in the coupling of KK graviton modes with the SM matter fields in comparison with the original RS model. Finally we found that the cross-section of SM processes involving virtual graviton exchange suffers a significant reduction in our model as compared to the RS model. At the same time, the decay-width for the decay of the lightest KK graviton modes into SM particles (like - dilepton or diphoton decay channels) will have a significant uplift in comparison with the RS model. As a whole, there is a significant observable modification in both production and decay signatures of KK graviton modes for the higher curvature model and it can be probed in some precision collider experiments in the foreseeable future. Finally, we showed the convergence of EFT for the allowed parameter space and validated our obtained results for the modifications in KK graviton masses and couplings with SM fields using the naive dimensional analysis. We also showed using the same framework that the higher-order curvature operators have subdominant corrections to our obtained results.\\

\noindent There are several interesting directions in which the present work can be extended. These include non-flat brane $f(\mathcal R)$ braneworld models~\cite{71,72} and multi-brane extensions~\cite{73,74}. It would also be worthwhile to investigate the radion sector and the corresponding radion KK modes within the present perturbative framework. Furthermore, a dedicated collider analysis could provide a more quantitative assessment of the phenomenological implications of the model. Some of these directions will be pursued in future work.

\section{Acknowledgments}

AK is supported through INSPIRE-SHE Scholarship by the Department of Science and Technology (DST), Government of India. IM is supported through Indian Association for the Cultivation of Science (IACS) Masters Fellowship. AK acknowledges Prof. Sourov Roy for some illuminating suggestions regarding the KK graviton phenomenology. AK also thanks Gahan Chattopadhyay for many useful discussions and suggestions. The authors would also like to thank the reviewer(s) for helping to improve the manuscript.

\section*{Data Availability Statement}

Being a theoretical study, no experimental data is associated with this work.

\section*{Code Availability Statement}

Apart from plotting the figures, no major code is associated with this work. This shall be made available on reasonable request.

\appendix

\section{Features of KK Gravitons: An Alternative Path Integral Approach}\label{A}

In this section, we revisit the well-known Randall-Sundrum KK graviton problem with pure Einstein gravity in the bulk from the perspective of Euclidean path integrals. This provides an alternative route to recover the standard RS results while offering several additional advantages. First, the formalism naturally yields the propagator corresponding to the RS Hamiltonian, which serves as a fundamental object for computing several physically relevant quantities. As an illustration, we show that the momentum-space Green's function can be constructed directly from the propagator, leading to the well-known correction to the four-dimensional Newtonian gravitational potential in the RS model. Secondly, the path-integral approach provides a complementary formulation of the KK graviton problem, thereby enriching the theoretical understanding of the model. It is worth emphasizing that, although the graviton equation assumes the form of a Schr\"odinger-like equation, the underlying problem remains entirely classical, with the quantum-mechanical analogy arising purely from its mathematical structure. \\

\noindent For $\xi > 0$, from eq.(\ref{3.4}) we read off the effective Hamiltonian of the system as follows
\begin{equation}\label{A1}
\hat{H} = -\partial_\rho^2 + \frac{15}{4\rho^2} \equiv \hat{p}^2 + V(\rho)
\end{equation}
where, $\hat p$ and $V(\rho)$ are the corresponding momentum operator and the potential in $r$ coordinate. Note that $\rho \equiv \xi + 1/k$ for $\xi >0$. Now the Euclidean propagator written in eq.(\ref{3.9}) can be written as follows
\begin{equation}\label{A2}
\mathcal K_E(\rho_f, \rho_i; T) = \lim_{\substack{\epsilon \to 0 \\ N \to \infty}} \int_0^\infty \cdots \int_0^\infty \left(\prod_{j=1}^{N-1} d\rho_j\right) \prod_{n=1}^N \left\langle \rho_n \left| e^{-\epsilon \hat{H}} \right| \rho_{n-1} \right\rangle
\end{equation}
where, we have discretized the total Euclidean time $\tau$ in $N$ number of intervals in steps of $\epsilon$. One can explicitly compute the matrix elements in eq.(\ref{A2}) and will get the following expression of the Euclidean propagator
\begin{equation}\label{A3}
\mathcal K_E(\rho_f, \rho_i; T) = \lim_{\substack{\epsilon \to 0 \\ N \to \infty}} \int_0^\infty \cdots \int_0^\infty \left(\prod_{j=1}^{N-1} d\rho_j\right) \prod_{n=1}^N \left[ \sqrt{\frac{\rho_n \rho_{n-1}}{2\epsilon}} \exp\left( -\frac{\rho_n^2 + \rho_{n-1}^2}{4\epsilon} \right) I_2\left(\frac{\rho_n \rho_{n-1}}{2\epsilon}\right) \right]
\end{equation}
where, $I_2$ is the modified Bessel's function of order $2$. For deriving the above expression for the propagator, one may need the following asymptotic expansion of the modified Bessel function of any general order $I_\nu$
\begin{equation}\label{A4}
I_\nu(x) = \frac{e^x}{\sqrt{2\pi x}} \left[ 1 - \frac{4\nu^2 - 1}{8x} + \frac{(4\nu^2 - 1)(4\nu^2 - 9)}{2!(8x)^2} - \cdots \right] \;\;\;\;\;;\;\;\text{for }x\to\infty
\end{equation}
Now using the following standard identity involving modified Bessel's function
\begin{equation}\label{A5}
\int_0^\infty x \exp\left[ -\frac{a x^2}{2} \right] I_2(bx) I_2(cx) \, dx = \frac{1}{a} \exp\left[ \frac{b^2 + c^2}{2a} \right] I_2\left(\frac{bc}{a}\right)
\end{equation}
we can express the Euclidean propagator in the following closed form
\begin{equation}\label{A6}
\mathcal K_E(\rho_f, \rho_i; T) = \frac{\sqrt{\rho_f \rho_i}}{2T} \exp\left(- \frac{\rho_i^2 + \rho_f^2}{4T} \right) I_2\left(\frac{\rho_i \rho_f}{2T}\right)
\end{equation}
We now need to impose the boundary conditions mentioned in eq.(\ref{3.8}) in order to obtain the correct graviton wavefunctions in eq.(\ref{3.6}) and (\ref{3.10}) as well as the graviton spectrum in eq.(\ref{3.11}) from the Euclidean propagator in (\ref{A6}). The graviton wavefunctions and the spectrum can be extracted from the propagator using the standard relation involving the wavefunctions and eigenvalues
\begin{align}\label{A7}
\mathcal K_E(\rho_f, \rho_i; T) &= \sum_{n} \Psi_n(\rho_f)\Psi_n(\rho_i) e^{-m_n^2 T}
\end{align}
and for doing so again one needs another identity involving Bessel functions
\begin{equation}\label{A8}
\int_{0}^{\infty} k \, J_\nu(kx) J_\nu(kx') \, dk = \frac{1}{x} \delta(x - x')
\end{equation}
After applying the boundary conditions appropriately, we get the discrete form for the Euclidean propagator. We call this propagator as the \textit{RS Propagator} and it can be explicitly written in $\xi$ coordinate as
\begin{align}\label{A9}
\mathcal K_{RS}(&\xi_f, \xi_i; T) = \frac{2k}{(k\xi_f+1)^{3/2}(k\xi_i+1)^{3/2}} \nonumber \\
&+ \sum_{n=1}^{\infty} \frac{2k e^{-2kb} \sqrt{(k\xi_f+1)(k\xi_i+1)}}{[J_2(x_n)]^2} J_2\left[m_{n}\left(\xi_f + \frac{1}{k}\right)\right] J_2\left[m_{n}\left(\xi_i + \frac{1}{k}\right)\right] e^{-\frac{m_{n}^2}{2} T}
\end{align}
with $m_{n} = kx_ne^{-kb}$. The masses of the massive KK modes are exponentially suppressed, hence it will be enough if we consider only the low-lying massive modes in the above sum. For practical purposes, it is sufficient to consider only the first massive KK mode and it can show all relevant physics associated with the RS gravity. We can calculate the Green's function in momentum space by integrating the propagator (\ref{A9}) over $T$ considering only the zero mode and the first massive mode
\begin{equation}\label{A10}
G_E(p;\xi) = \int_0^\infty dT\;e^{-p^2T/2}\mathcal K_{RS}(\xi,\xi;T) \simeq \frac{4k}{(k\xi+1)^3 p^2} + \frac{4k e^{-2kb}(k\xi+1) \left[J_2\left(m_{1}\left(\xi+1/k\right)\right)\right]^2}{[J_2(x_1)]^2 \left(p^2 + m_{1}^2\right)}
\end{equation}
The above Green's function in momentum space is essential to calculate static gravitational potential for RS gravity. Doing a 3D Fourier transform of eq.(\ref{A10}) with $p^0 = 0$, we get the effective gravitational potential for a mass $M$ at some point $\xi$ in the extra dimension as follows
\begin{equation}\label{A11}
    V_{eff}(r) = -\frac{G_NM}{r}\left[1 + \mu\frac{e^{-m_{1}r}}{r}\right]
\end{equation}
where, $G_N$ is the Newton's gravitational constant and $\mu$ is called the Yukawa strength and it is given as a function of the extra dimensional coordinate $\xi$ by the following form
\begin{equation}\label{A12}
    \mu(\xi) = \frac{(k\xi+1)^4 e^{-2kb} \left[J_2\left(m_{1}\left(\xi + \frac{1}{k}\right)\right)\right]^2}{[J_2(x_1)]^2}
\end{equation}
and $r$ is the usual radial coordinate for expressing any central potential. We can see that at large-$r$, only the $1/r$ piece dominates and this is the expected long range nature of 4D gravitational potential. However, the Yukawa part $e^{-m_{1}r}/r$ is very short range in nature and this is the correction to the standard Newton's gravitational potential. As a whole, the potential in eq.(\ref{A11}) reproduces the 4D \textit{Einstein gravity} (or, the correction to the Newton's gravity).

\section{Various Constants and Coefficients}\label{B}

The coefficients appearing in eq.(\ref{4.21}) are the followings
\begin{align}
P_m &= \frac{2}{K^2}\left[-Kk \ln x_m + \frac{K(K - k)}{4} + 8Kk \delta_{m1}\right]\\
Q_m &= \frac{2}{K^2}\left[-\frac{K - k}{4}\ln x_m + \frac{k}{2}(\ln x_m)^2 + (2K - 2k)\delta_{m1} + 8k \delta_{m2}\right]
\end{align}
where,
\begin{align}
\delta_{m1} &= \int_1^{x_m} \frac{du}{u^3} J_2(u)\label{B3} \\ 
\delta_{m2} &= \int_1^{x_m} \frac{du}{u^3} J_2(u) 
\ln\!\left(\frac{u}{x_m}\right)\label{B4}
\end{align}
and $x_m$ are the zeros the Bessel function $J_1$, that is, $J_1(x_m) = 0$. \\

\noindent The coefficients appearing in eq.(\ref{4.24}) are the followings
\begin{align}
\mathcal C_{n0} &= \frac{2e^{-Kb}}{J_2(x_n)}
\left[
k\left(\frac{1}{2}-\mu_n\right)b^2
+ \left(\frac{C_{1n}}{K^2} - 2\mu_n - \frac{2k}{K}\nu_n\right)b
+ \left(\frac{C_{0n}}{K^2} - \frac{2}{K}\nu_n - \frac{k}{K^2}\rho_n\right)
\right] \\
\mathcal P_{nm} &= -\frac{2k\,\Phi_{nm}^{(w)}}{(x_n^2-x_m^2)J_2(x_m) J_2(x_n)} \\
\mathcal Q_{nm} &= \frac{16k\,\Phi_{nm} - 4K\,\Phi_{nm}^{(w)} - 4k\,\Psi_{nm}^{(w)}}{K (x_n^2-x_m^2)J_2(x_m) J_2(x_n)} \\
\mathcal L_{nm} &= \frac{4(K-k)\Phi_{nm} + 16k\Psi_{nm} - 4K\,\Psi_{nm}^{(w)} - 2k\,\Omega_{nm}^{(w)}}{K^2 (x_n^2-x_m^2)J_2(x_m) J_2(x_n)}
\end{align}
where,
\begin{align}
\mu_n &= \int_0^{x_n} \frac{du}{u} J_2(u)\\
\nu_n &= \int_0^{x_n} \frac{du}{u} J_2(u)\ln\frac{u}{x_n} \\
\rho_n &= \int_0^{x_n} \frac{du}{u} J_2(u)\left({\ln\frac{u}{x_n}}\right)^2 \\
C_{0n} &= -\frac{K-k}{4}\ln x_n + \frac{k}{2}(\ln x_n)^2 + (2K-2k)\delta_{n1} + 8k\,\delta_{n2}\\
C_{1n} &= -Kk\ln x_n + \frac{K(K-k)}{4} + 8Kk\,\delta_{n1}
\end{align}
\begin{align}
\Phi_{nm} &= \int_0^{x_n} \frac{du}{u} J_2\!\left(\frac{x_m}{x_n}u\right) J_2(u) \\
\Psi_{nm} &= \int_0^{x_n} \frac{du}{u} J_2\!\left(\frac{x_m}{x_n}u\right) J_2(u)\, \ln\frac{u}{x_n}\\
\Phi_{nm}^{(w)} &= \int_0^{x_n} du\, u\, J_2\!\left(\frac{x_m}{x_n}u\right) J_2(u) \\
\Psi_{nm}^{(w)} &= \int_0^{x_n} du\, u\, J_2\!\left(\frac{x_m}{x_n}u\right) J_2(u)\, \ln\frac{u}{x_n} \\
\Omega_{nm}^{(w)} &= \int_0^{x_n} du\, u\, J_2\!\left(\frac{x_m}{x_n}u\right) J_2(u)\, {\left(\ln\frac{u}{x_n}\right)}^2
\end{align}
and $\delta_{n1},\delta_{n2}$ are defined in (\ref{B3}) and (\ref{B4}), respectively.

\end{document}